\documentclass[prd,showpacs,preprintnumbers,twocolumn,amsmath,nofootinbib,amssymb,floatfix]{revtex4}
\usepackage{graphicx,color,dcolumn,booktabs,bm}
\usepackage{epstopdf}
\usepackage{longtable,lscape}
\usepackage{txfonts}
\usepackage{overpic}
\usepackage{float}
\usepackage{amssymb}
\usepackage{amsmath}
\usepackage{epstopdf}
\usepackage{indentfirst}
\usepackage{feynmf}  
\usepackage{slashed}  
\usepackage{cases}
\usepackage{ulem}
\usepackage{color}
\usepackage{float}
\usepackage{multirow}
\usepackage{tikz}
\usepackage{epstopdf}
\usepackage{epsfig,dsfont,amssymb,amsmath,amsfonts,amsbsy,mathrsfs}
\usepackage{multirow}
\usepackage{graphicx}  
\usepackage{float}  
\usepackage{subfigure}  
\usepackage{array}
\usepackage{microtype}

\usepackage[colorlinks, citecolor=blue,anchorcolor=red,menucolor=red, linkcolor=red,filecolor=red,runcolor=red,urlcolor=blue,frenchlinks=true]{hyperref}
\makeatletter
\@addtoreset{equation}{section}
\makeatother

\newcommand{\nc}{\newcommand}
\nc{\tj}[1]{\textcolor{red}{Tianjin: #1}}

\usepackage{soul}
\allowdisplaybreaks
\begin{document}
\title{New states $X(1910)$ and $X(2300)$ and higher light excited $J^{PC}=1^{+-}$ mesons}
 \author{Ya-Rong Wang$^{1,2}$}
 \author{Xiao-Hai Liu$^{1}$ }\email{ xiaohai.liu@tju.edu.cn}
\author{Cheng-Qun Pang$^{3,4,5}$  }\email{xuehua45@163.com}
 \author{Hao Chen$^{4}$  }
\affiliation{ $^{1}$Center for Joint Quantum Studies and Department of Physics, School of Science, Tianjin University, Tianjin 300350, China\\$^2$Center for Theoretical Physics, School of Physics and Optoelectronic Engineering, Hainan University, Haikou 570228, China\\
$^3$School of Physics Optoelectronic Engineering, Ludong University, Yantai 264000, China\\
$^4$College of Physics and Electronic Information Engineering, Qinghai Normal University, Xining 810000, China
\\$^5$Lanzhou Center for Theoretical Physics, Key Laboratory of Quantum Theory and Applications of MoE, and Key Laboratory of Theoretical Physics of Gansu Province, Lanzhou University, Lanzhou, Gansu 730000, China
}

\date{\today}

\begin{abstract}

The BESIII Collaboration recently reported the observation of two new resonances, $X(1910)$ and $X(2300)$,  which have sparked our interest in studying the light hadron family with  $J^{PC}=1^{+-}$ . In this work, we investigate the mass spectra and Okubo-Zweig-Iizuka (OZI)-allowed two-body strong decays of $b_1$, $h_1$, and $h_1^\prime$ using the modified Godfrey-Isgur (MGI) quark model and quark-pair creation (QPC) model with newly fitted parameters. We also explore the possibility of identifying $X(1910)$ and $X(2300)$ as $h_1$ or $h_1^\prime$ states. 
Our numerical results suggest that $X(1910)$ could be a promising candidate for the  $h_1^\prime(2^1P_1)$ state with quark content $s\bar{s}$, while the structure of $X(2300)$ remains uncertain.

\end{abstract}
\maketitle

\section{introduction}\label{sec1}

Very recently, the BESIII Collaboration announced a new axial-vector state [denoted as $X(2300)$] around 2.3 GeV in the $\phi\eta$ and $\phi\eta^\prime$ invariant mass spectra. Its significance is $9.6\sigma$, and the corresponding mass and width are determined to be $2316 \pm 9\pm30$ MeV and $89 \pm 15\pm26$ MeV by a scan of the log likelihood value \cite{BESIII:2024nhv}. Prior to this, the BESIII Collaboration also announced the observation of the $X(1910)$ state in the $J/\psi \to \phi\pi^0\eta$ process, with $J^{PC} = 1^{+-}$, mass M$ = 1911 \pm 6 \pm 14$ MeV, and width $\Gamma = 149 \pm 12 \pm 23$ MeV at the BEPCII collider \cite{BESIII:2023zwx}. Its statistical significance is $24.0\sigma$. This discovery intrigued us to investigate whether $X(2300)$ and $X(1910)$ could belong to the axial vector mesons.
\par
The $J^{PC}=1^{+-}$ meson families of low energy are emerging as a powerful source of information on hadron dynamics in quark models \cite{Steph:1985ff}. Subsequent studies in the context of QCD and large $N_c$ behavior, combined with phenomenology, have shown that the vector mesons are largely $q\bar{q}$ objects \cite{PELAEZ20161, PhysRevD.99.094020}. 
There have already been lots of works for the low-lying $J^{PC}=1^{+-}$ mesons. Godfrey and Isgur calculated the mass spectrum using the relativized quark model (GI model) \cite{Godfrey:1985xj}. Ebert calculated masses of the ground state and the orbitally and radially excited states of quark-antiquark mesons composed of light ($u$, $d$, $s$) quarks within the framework of the relativistic quark model based on the quasipotential approach, and the result showed $b_1(1960)$ is the second radially excited state of $b_1(1235)$ \cite{Ebert:2009ub}. 
In 2015, the Lanzhou group systematically studied the mass spectrum of the axial-vector meson family by Regge trajectory analysis and calculated two-body Okubo-Zweig-Iizuka (OZI)-allowed strong decays. At the same time, the quark-pair creation (QPC) model with simple harmonic oscillator wave function was adopted to calculate their OZI-allowed strong decay width \cite{Chen:2015iqa}.

Samson {\it{et al}}. investigated the isoscalar axial-vector $h_1$ mesons using a coupled-channel formalism, solved the off-shell coupled integral equations, and discussed the dynamical generation of the $h_1(1170)$ \cite{Clymton:2024pql}. Their analysis explained why some experiments observed only the lower pole \cite{Clymton:2024pql}.
The mass of the $h_1^\prime(^1P_1)$ state, predominantly of $s\bar{s}$ content, is approximately $1495.18 \pm 8.82$ MeV, according to Li {\it{et al}}. \cite{Li:2005eq}.
Nils made numerous predictions regarding $1^{++}$ and $1^{+-}$ mesons, concluding that the mass and width of the $b_1(1235)$ are 1242 MeV and 114 MeV, respectively \cite{Tornqvist:1981bs}.
Barnes {\it{et al}}. qualitatively compared the reported total width with expectations for a $1^1P_1$ $s\bar{s}$ state by varying the assumed $h_1$ mass, ultimately concluding that the assignment of $h_1(1380)$ to $1^1P_1$ $s\bar{s}$ appears plausible \cite{Barnes:2002mu}. Meson flavor mixing has been extensively investigated \cite{Cheng:2011fk, Cheng:2013cwa, Liu:2014doa, Cheng:2011pb, Li:2005eq, Dudek:2011tt}.
Furthermore, the mass and width of the $h_1(1170)$ are 1177 MeV and 352 MeV, respectively \cite{Tornqvist:1981bs}. 
Moreover, the $h_1(1830)$, observed in the BESIII data for $J/\psi \to \eta{K^{*0}}\bar{K}^{*0}$, is considered as a $K^*K^*$ molecule \cite{Albaladejo:2016raz}, and will be not discussed in this work.
\par
In 2019, the BESIII Collaboration announced the observation of $X(2100)$ in the $\phi\eta^\prime$ invariant mass spectrum of the $J/\psi \to \phi\eta\eta^\prime$ decay \cite{Ablikim:2018xuz}. 
The Lanzhou group has found that it is suitable to explain $X(2100)$ to be as the second radially excited state of $h_1(1380)$ since the experimental width of $X(2100)$ can be reproduced and the branching ratio of $X(2100) \to \phi\eta^\prime$ has a sizable contribution to the total width \cite{Wang:2019qyy} .   
The experimental information of the observed $J^{PC}=1^{+-}$ mesons is listed in Table \ref{massinpdg}. Nevertheless, could the newly discovered particles $X(2100)$ and $X(2300)$ be the same resonant state? 
Additionally, does particle $X(1910)$ also belong to the same family as them? Therefore, we will study $J^{PC}= 1^{+-}$ mesons $b_1$, $h_1$, as well as $h_1^\prime$. 
\par

In this work,  
we adopt the quark flavor mixing scheme to study $h_1-{h_1^\prime}$ meson mixing and give a discussion about the total decay widths of $h_1$ and $h_1^\prime$ mesons depending on mixing angle. 
We will study $J^{PC}= 1^{+-}$ mesons $b_1$, $h_1$, as well as $h_1^\prime$ and compare the results with experimental values. Furthermore, the possibility of treating the new states $X(1910)$ and $X(2300)$ as $h_1$ or $h_1^\prime$ states will be discussed.
First, we will adopt the Regge trajectory to help us decide the approximate mass range of $X(1910)$ and $X(2300)$ as $h_1$ or $h_1^\prime$ states as well as other $J^{PC}= 1^{+-}$ mesons, which lays a good foundation for the next step. Then, we will study the mass spectra and the strong decay behaviors of $b_1$, $h_1$, and $h_1^\prime$ states with the modified Godfrey-Isgur (MGI)  quark model \cite{Song:2015nia, Pang:2018gcn} and the $^3P_0$ model.

\par
The paper is organized as follows: In Sec. \ref{2}, we briefly reviewed the Regge trajectory, the MGI model, and the $^3P_0$ model.
In Sec. \ref{sec3}, the obtained results for the masses of $b_1$, $h_1$, and $h_1^\prime$ are presented and compared with available experimental data and other results from different methods. Then, we systematically study the OZI-allowed two-body strong decay behaviors of the newly observed $X(1910)$, $X(2300)$, $b_1$, $h_1$, and $h_1^\prime$ states. Finally, a conclusion is given in Sec. \ref{sec4}.      
\begin{table}[htbp]
\renewcommand{\arraystretch}{1.5}
\centering
\caption{Resonance parameters of the $h_1$, $h_1^\prime$, and $b_1$ states collected in the PDG. The unit of the mass and width is MeV.
 \label{massinpdg}}
\[\begin{array}{ccc}
\hline\hline
\text{State}&\text{Mass}     &\text{Width}       \\\midrule[1pt]
 h_1(1170)$~\cite{ParticleDataGroup:2024cfk}$   &1166\pm6        &375\pm35    \\
 h_1(1415)$~\cite{ParticleDataGroup:2024cfk}$   &1409^{+9}_{-8}  &78\pm11      \\
 h_1(1595)$~\cite{BNL-E852:2000poa}$   &1594^{+18}_{-60}&384^{+90}_{-120} \\
 h_1(1965)$~\cite{Anisovich:2002xoo}$   &1965\pm45       &345\pm75  \\
 h_1(2215)$~\cite{Anisovich:2002xoo}$   &2215\pm40       &325\pm55    \\
 b_1(1235)$~\cite{ParticleDataGroup:2024cfk}$   &1229.5\pm3.2    &142\pm9     \\
 b_1(1960)$~\cite{Anisovich:2002xoo}$   &1960\pm35       &230\pm50    \\
 b_1(2240)$~\cite{Anisovich:2002xoo}$   &2240\pm35       &320\pm85   \\
 \hline\hline
\end{array}\]
\end{table}

 \section{Models employed in this work} \label{2}
 \subsection{Regge trajectory}
\begin{figure*}[htbp]
\centering%
\includegraphics[scale=1.1]{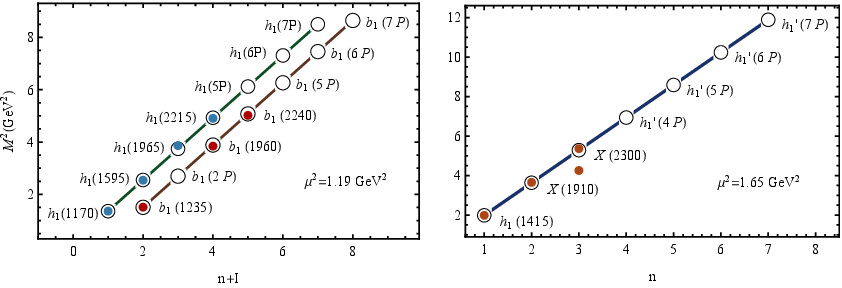}
\put(-136,59){\fontsize{7.5pt}{7pt}\selectfont $X(2100)$}
\caption{The Regge trajectories of $h_1$, $b_1$, and $h_1^\prime$ 
families. Here the open circle and the filled geometry are the theoretical and experimental values, respectively. Here, n is the radial quantum number of the corresponding meson and I is the isospin of mesons.}
\label{regge}
\end{figure*}

It is efficient to investigate a light-meson spectrum by determining the Regge trajectory \cite{Chew:1962eu,Anisovich:2000kxa}.
The following relation is satisfied by the masses and radial quantum numbers of the light mesons that belong to the same meson family
\begin{eqnarray}
M^2=M_0^2+(n-1)\mu^2, \label{rt}
\end{eqnarray}
where {$M_0$} is the mass of the ground state, 
$n$ represents the radial quantum number of the corresponding meson with  the mass $M$, and $\mu^2$ is the trajectory slope. We take $\mu^2=1.65$ $\text{GeV}^2$ for $h_1^\prime$ family and 1.19 $\text{GeV}^2$ for $h_1$ and $b_1$ family as shown in Fig. \ref{regge}.
\par

Although Regge trajectories may not adhere strictly to the function of particle spin, they can still be considered valid approximations. This is because, for each quantum number, only one or two excitations are typically known.
Subsequently, we will calculate the mass spectra using the MGI model, which obviously takes into account spin breaking.

\par

 \subsection{The modified GI model}

{\color{black}
The MGI model was proposed in Ref. \cite{Song:2015nia} based on GI model \cite{Godfrey:1985xj}.
Subsequently, great success has been achieved in light hadron spectroscopy \cite{Pang:2018gcn, Wang:2022xxi, Wang:2021abg, Wang:2024yvo, Feng:2022hwq}. In the MGI model, the interaction between quark and antiquark is

\begin{equation}\label{hh}
\widetilde{H}=\sqrt{m_1^2+\mathbf{p}^2}+\sqrt{m_2^2+\mathbf{p}^2}+\widetilde{V}^{\mathrm{eff}},
\end{equation}
where $m_1$ and $m_2$ denote the masses of the quark and antiquark, respectively, and the effective potential of the $q\bar{q}$ interaction $\widetilde{V}^{\mathrm{eff}}$ includes short-range $\gamma^{\mu}\otimes\gamma_{\mu}$ one-gluon-exchange interaction and $1\otimes1$ linear confinement interaction.
Specifically, the effective potential is given by

\begin{equation}\label{V}
\widetilde{V}^{\mathrm{eff}}=\widetilde{G}_{12}+\widetilde{V}^{\mathrm{cont}}+\widetilde{V}^{\mathrm{tens}}+\widetilde{V}^{\mathrm{so(v)}}+\widetilde{S}_{12}(r)+\widetilde{V}^{\mathrm{so(s)}},
\end{equation}
which includes the Coulomb term($\widetilde{G}_{12}$), the contact term($\widetilde{V}^{\mathrm{cont}}$), the tensor ($\widetilde{V}^{\mathrm{tens}}$), vector spin-orbit term ($\widetilde{V}^{\mathrm{so(v)}}$), screened confinement term [$\widetilde{S}_{12}(r)$], and scalar spin-orbit interaction term ($\widetilde{V}^{\mathrm{so(s)}}$).

The spin-independent terms of the norelativistic potential read

\begin{equation}
\begin{aligned}
\tilde{G}(r)=-\sum_k\frac{4\alpha_k}{3r}\left[\frac{2}{\sqrt{\pi}}\int_0^{\gamma_{k}r} e^{-x^2}dx\right],\\
\end{aligned}
\end{equation}

where $\alpha_k=(0.25,0.15,0.2)$ and $\gamma_k=(1/2,\sqrt{10}/2,\sqrt{1000}/2)$ for $k=1,2,3$ \cite{Godfrey:1985xj}, 
and

\begin{equation}
{S}(r)=\frac{b(1-e^{-\mu r})}{\mu}+c,
\end{equation}

{where $\mu$ is} screened parameter whose particular value is needed to be fixed by the comparisons between theory and experiment.
\par
The effective potential $\widetilde{V}^{\mathrm{eff}}$ considered
relativistic effects, particularly in meson systems, which are embedded in two ways.
First, a smearing function for a meson $q\bar{q}$ is introduced, using nonlocal interactions and new $\mathbf{r}$ dependencies, which has the form

\begin{equation}
\rho_{ij} \left(\mathbf{r}-\mathbf{r'}\right)=\frac{\sigma_{ij}^3}{\pi ^{3/2}}e^{-\sigma_{ij}^2\left(\mathbf{r}-\mathbf{r'}\right)^2},
\end{equation}
with
\begin{align}
   \sigma_{ij}^2=\sigma_0^2\Bigg[\frac{1}{2}+\frac{1}{2}\left(\frac{4m_im_j}{(m_i+m_j)^2}\right)^4\Bigg]+
  s^2\left(\frac{2m_im_j}{m_i+m_j}\right)^2,
\end{align}
where the values of $\sigma_0$ and $s$ are defined in Table \ref{MGI}.
The Coulomb term $\widetilde{G}_{12}(r)$ is spin independent and is defined as
\begin{equation}
\begin{split}
\widetilde{G}_{ij}(r)=&\int d^3{\bf r}^\prime \rho_{ij}({\bf r}-{\bf r}^\prime)G(r^\prime)
=\sum\limits_k-\frac{4\alpha_k }{3r}{\rm erf}(\tau_{kij}r),
\end{split}
\end{equation}
the values of $\tau_{kij}$ could be obtained by
\begin{equation}
\tau_{kij}=\frac{1}{\sqrt{\frac{1}{\sigma_{ij}^2}+\frac{1}{\gamma_k^2}}}.
\end{equation}
We express the potential $\widetilde{S}_{12}(r)$ as
\begin{eqnarray}
\widetilde{S}_{12}(r)&=& \int d^3 {\bf{r}}^\prime
\rho_{12} ({\bf r}-{\bf r}^\prime)S(r^\prime)\nonumber\\
&=& \frac{b}{\mu r}\Bigg[r+e^{\frac{\mu^2}{4 \sigma^2}+\mu r}\frac{\mu+2r\sigma^2}{2\sigma^2}\Bigg(\frac{1}{\sqrt{\pi}}
\int_0^{\frac{\mu+2r\sigma^2}{2\sigma}}e^{-x^2}dx-\frac{1}{2}\Bigg) \nonumber\\
&&-e^{\frac{\mu^2}{4 \sigma^2}-\mu r}\frac{\mu-2r\sigma^2}{2\sigma^2}\Bigg(\frac{1}{\sqrt{\pi}}
\int_0^{\frac{\mu-2r\sigma^2}{2\sigma}}e^{-x^2}dx-\frac{1}{2}\Bigg)\Bigg]  \nonumber \\
&&+c. \nonumber\label{Eq:pot}
\end{eqnarray}
Second, due to relativistic effects, the general potential should depend on the mass center of the interacting quarks. Momentum-dependent factors, which are unity in the nonrelativistic limit, are applied as
\begin{equation}
\tilde{G}_{12}(r)\to \tilde{G}_{12}=\left(1+\frac{\mathbf{p}^2}{E_1E_2}\right)^{1/2}\tilde{G}_{12}(r)\left(1 +\frac{\mathbf{p}^2}{E_1E_2}\right)^{1/2}.
\end{equation}
For the spin-dependent terms, the semirelativistic correction could be written as

\begin{equation}
\label{vsoij}
  \tilde{V}^i_{\alpha \beta}(r)\to\tilde{V}^i_{\alpha \beta}= \left(\frac{m_\alpha m_\beta}{E_\alpha E_\beta}\right)^{1/2+\epsilon_i} \tilde{V}^i_{\alpha \beta}(r)\left(\frac{m_\alpha m_\beta}{E_\alpha E_\beta}\right)^{1/2+\epsilon_i},
\end{equation}
where $\tilde{V}^i_{\alpha \beta}(r)$ delegate the contact term, tensor term, vector term, and scalar spin-orbit term, and $\epsilon_i=\epsilon_c$, $\epsilon_t$, $\epsilon_{\rm so(v)}$, and $\epsilon_{\rm so(s)}$ impacts the potentials $\widetilde{V}^{\mathrm{cont}}$, $\widetilde{V}^{\mathrm{tens}}$, $\widetilde{V}^{\mathrm{so(v)}}$, and $\widetilde{V}^{\mathrm{so(s)}}$, respectively \cite{Wang:2021abg}.
They have the following form
\begin{equation}\label{Vcont}
\begin{split}
\widetilde{V}^{\mathrm{cont}}=\frac{2{\bf S}_1\cdot{\bf S}_2}{3m_1m_2}\nabla^2\widetilde{G}_{12}^c,
\end{split}
\end{equation}
\begin{equation}\label{Vtens}
\begin{split}
\widetilde{V}^{\mathrm{tens}}=&-\left(\frac{3{\bf S}_1\cdot{\bf r}{\bf S}_2\cdot{\bf r}/r^2-{\bf S}_1\cdot{\bf S}_2}{3m_1m_2}\right)\left(\frac{\partial^2}{\partial r^2}-\frac{1}{r}\frac{1}{\partial r}\right)\widetilde{G}_{12}^t,
\end{split}
\end{equation}
\begin{equation}\label{Vsov}
\begin{split}
\widetilde{V}^{\mathrm{so(v)}}=&\frac{{\bf S}_1\cdot {\bf L}}{2m_1^2}\frac{1}{r}\frac{\partial\widetilde{G}_{11}^{\rm so(v)}}{\partial r}+\frac{{\bf S}_2\cdot {\bf L}}{2m_2^2}\frac{1}{r}\frac{\partial\widetilde{G}_{22}^{\rm so(v)}}{\partial r}
\\
 &
+\frac{({\bf S}_1+{\bf S}_2)\cdot {\bf L}}{m_1m_2}\frac{1}{r}\frac{\partial\widetilde{G}_{12}^{\rm so(v)}}{\partial r},\\
\end{split}
\end{equation}

\begin{equation}\label{Vsos}
\begin{split}
\widetilde{V}^{\mathrm{so(s)}}=&-\frac{{\bf S}_1\cdot {\bf L}}{2m_1^2}\frac{1}{r}\frac{\partial\widetilde{S}_{11}^{\rm so(s)}}{\partial r}-\frac{{\bf S}_2\cdot {\bf L}}{2m_2^2}\frac{1}{r}\frac{\partial\widetilde{S}_{22}^{\rm so(s)}}{\partial r}.\\
\end{split}
\end{equation}
For example, base on the rule as shown in Eq. (\ref{vsoij}) has the form

\begin{equation}\label{Vcont}
\begin{split}
\widetilde{V}^{\mathrm{cont}}=&\left(\frac{m_1 m_2}{E_1 E_2}\right)^{1/2+\epsilon_c}\frac{2{\bf S}_1\cdot{\bf S}_2}{3m_1m_2}\nabla^2\widetilde{G}_{12}(r)\left(\frac{m_1 m_2}{E_1 E_2}\right)^{1/2+\epsilon_c}\\
=&\left(\frac{m_1 m_2}{E_1 E_2}\right)^{1/2+\epsilon_c}\frac{2{\bf S}_1\cdot{\bf S}_2}{3m_1m_2}\nabla^2\left(-\sum\limits_k\frac{4\alpha_k }{3r}{\rm erf}(\tau_{k12}r)\right)\\
&\left(\frac{m_1 m_2}{E_1 E_2}\right)^{1/2+\epsilon_c}. \nonumber
\end{split}
\end{equation}

Through diagonalizing the Hamiltonian matrix as depicted in Eq. (\ref{hh}) under a simple harmonic oscillator (SHO) base, the mass and the spatial wave function of the meson can be obtained and applied to the strong decay process.

}

 \subsection{A brief review of the \text{$^3P_0$} model}

The $^3P_0$ model (also called QPC  model) was initially put up by Micu \cite{Micu:1968mk} and further developed by the Orsay group \cite{LeYaouanc:1972ae, LeYaouanc:1973xz, LeYaouanc:1974mr, LeYaouanc:1977gm, LeYaouanc:1977ux}. This model is widely applied to the OZI-allowed two-body strong decays of mesons in Refs. \cite{vanBeveren:1982qb, Titov:1995si, Ackleh:1996yt, Blundell:1996as, Bonnaz:2001aj, Zhou:2004mw, Lu:2006ry, Zhang:2006yj, Luo:2009wu, Sun:2009tg, Liu:2009fe, Sun:2010pg, Rijken:2010zza, Ye:2012gu, Wang:2012wa, He:2013ttg, Sun:2013qca,  Wang:2014sea, Chen:2015iqa, Pang:2018gcn, Wang:2022juf, Wang:2022xxi, Li:2022khh, Li:2022bre, Wang:2020due, Pang:2019ttv, Pang:2017dlw, Pang:2015eha,  Pang:2014laa, Wang:2024yvo, Wang:2021abg, feng:2021igh, Feng:2022hwq}.
The transition operator $\mathcal{T}$ describes a quark-antiquark pair (denoted by indices $3$ and $4$) creation from vacuum that
$J^{PC}=0^{++}$.
For the process $A\to B+C$, $\mathcal{T}$ can be written as \cite{Wang:2019jch}
{\begin{align}\label{gamma}
\mathcal{T} = & -3\gamma \sum_{m}\langle 1m;1~-m|00\rangle\int d \mathbf{p}_3d\mathbf{p}_4\delta ^3 (\mathbf{p}_3+\mathbf{p}_4) \nonumber \\
 & ~\times \mathcal{Y}_{1m}\left(\frac{\textbf{p}_3-\mathbf{p}_4}{2}\right)\chi _{1,-m}^{34}\phi _{0}^{34}
\left(\omega_{0}^{34}\right)_{ij}b_{3i}^{\dag}(\mathbf{p}_3)d_{4j}^{\dag}(\mathbf{p}_4),
\end{align}}
where $\mathcal{Y}_l^m(\vec{p})\equiv$ $p^lY_l^m(\theta_p,\phi_p)$ is the solid harmonics.$\chi$, $\phi$, and $\omega$ denote the spin, flavor, and color wave functions respectively. Subindices $i$ and $j$ are the color index of the $q\bar{q}$ pair. The amplitude ${M}^{{M}_{J_{A}}M_{J_{B}}M_{J_{C}}}$ of the decay process is defined with the transition operator $\mathcal{T}$,
\begin{eqnarray}
\langle BC|\mathcal{T}|A \rangle = \delta ^3(\mathbf{P}_B+\mathbf{P}_C)\mathcal{M}^{{M}_{J_{A}}M_{J_{B}}M_{J_{C}}}.
\end{eqnarray}
Finally, the general form of the decay width can be expressed as
\begin{eqnarray}
\Gamma&=&\frac{\pi}{4} \frac{|\mathbf{P}|}{m_A^2}\sum_{J,L}|\mathcal{M}^{JL}(\mathbf{P})|^2,
\end{eqnarray}
where $m_{A}$ is the mass of the initial state $A$, and
$\bf{P}$ is the three-momentum of meson B in the rest frame of meson A. The two decay amplitudes can be related by the Jacob-Wick formula \cite{Jacob:1959at}
\begin{equation}
\begin{aligned}
\mathcal{M}^{JL}(\mathbf{P}) = &\frac{\sqrt{4\pi(2L+1)}}{2J_A+1}\sum_{M_{J_B}M_{J_C}}\langle L0;JM_{J_A}|J_AM_{J_A}\rangle \\
    &\times \langle J_BM_{J_B};J_CM_{J_C}|{J_A}M_{J_A}\rangle \mathcal{M}^{M_{J_{A}}M_{J_B}M_{J_C}}.
\end{aligned}	
\end{equation}
A dimensionless parameter $\gamma$ in this model depicts the strength of the creation of $q\bar{q}$ from vacuum.  
The $\gamma$ value  depends on the reduced mass of quark-antiquark in the decaying meson, i.e. the $\gamma$ value of a meson with $n\bar{n}$ quark component is larger than  that of a  meson with $s\bar{s}$ component  \cite{Pang:2019ttv, Segovia:2012cd, Pang:2018gcn, Xiao:2018iez}.  
In this work, we take the $\gamma$ value to be $10.16$ for $n\bar{n}$ states, which fit the total width of  all $n\bar{n}$ states, and $7.20$ for the $s\bar{s}$ states, which fit the total width of the $s\bar{s}$ states listed in Table \ref{massinpdg}.
In our calculation, the spatial wave functions of the discussed $J^{PC}=1^{+-}$ meson family are obtained by the MGI model discussed in the preceding section.

We have adjusted the parameter of the MGI model by fitting with the experimental data based on previous work \cite{Pang:2018gcn} as shown in Table \ref{MGI}. The particles $b_1$, $h_1$, $a_1$, $a_2$, $f_1$, $f_2$, $K_1$, and ${K_1}^\prime$ correspond to the resonances $b_1(1235)$, $h_1(1170)$, $a_1(1260)$, $a_2(1320)$, $f_1(1285)$, $f_2(1270)$, $K_1(1270)$, and ${K_1}(1400)$, respectively. Other decay modes, such as $\omega(2D)$, $\omega(3D)$, $K^*(3S)$, $\phi(1D)$, $\rho(4S)$, and so on, represent the states that exist theoretically but have not yet been observed experimentally in Tables \ref{1pdecay} to  \ref{7pdecay}, and their mass is taken from the MGI model.  
Some decay channels with widths less than 1 MeV are omitted.  
\begin{table}[htbp]
\renewcommand{\arraystretch}{1.5}
\caption{Parameters and their values in this work. \label{MGI}}
\begin{center}
\begin{tabular}{cccc}
\hline\hline
Parameter &  value &Parameter &  value  \\
 \midrule[1pt]
$m_u$ (GeV)       &0.162    &{$\sigma_0$ (GeV)}   &{1.791}\\
$m_d$ (GeV)       &0.162    &{$s$ }          &{0.711}\\
$m_s$ (GeV)       &0.377    &$\mu$ (GeV)          &0.0779 \\
$b$ (GeV$^2$)     &0.222    &$c$ (GeV)            &$-0.228$\\
$\epsilon_c$      &-0.137   &$\epsilon_{so(v)}$     &0.0550\\
$\epsilon_{so(s)}$  &0.366    &$\epsilon_t$         &0.493\\
\hline\hline
\end{tabular}
\end{center}
\end{table}

\section{Numerical results and phenomenological analysis}\label{sec3}
We plot the Regge trajectory of $h_1^\prime$, $h_1$, and $b_1$ in Fig. \ref{regge}. By analyzing the Regge trajectory, the results demonstrate that $h_1(1170)$, $h_1(1595)$, $h_1(1965)$, and $h_1(2215)$ are $1P$, $2P$, $3P$, and $4P$ states, respectively. 
The $b_1(1960)$ and the $b_1(2240)$ are excited states of $b_1(1235)$. 
The situation for the $h_1^\prime$ is less straightforward than expected. The new observed state $X(1910)$ can be a good candidate of the first radially excited state of $h_1(1415)$. 
The situation of $h_1^\prime(3P)$ is still misty. 
The mass of $X(2100)$ is lower than the theoretical result of the Regge trajectory, at the same time, the new state $X(2300)$ can be considered as the $h_1^\prime(3P)$ state if only considered the Regge trajectory. This situation prompted us to continue to investigate its decay behavior.
\par
 Applying the MGI model with the new parameters in Table \ref{MGI}, the mass spectra of $J^{PC}=1^{+-}$ states are shown in the Table \ref{mass}. Ishida {\it{et al.}} have extended the covariant oscillator quark model (COQM) so as to include one-gluon exchange (OGE) effects covariantly and applied it to investigate the light-quark meson spectra \cite{PhysRevD.35.265}. 
 Ebert {\it{et al}}. calculated the mass spectra of light quark-antiquark mesons in the framework of the QCD-moticated relativistic quark model, which were carried out without application of the unjustified nonrelativistic $v/c$ expansion \cite{Ebert:2009ub}.
Li {\it{et al.}} have calculated the $s\bar{s}$ spectrum up to the mass range of $\sim 2.7$ GeV with a nonrelativistic linear quark potential model, where the model parameters were partially adopted from a calculation of the $\Omega$ spectrum \cite{Li_2021}.
Vijande {\it{et al.}}  have performed an exhaustive study of the meson spectra from the light $n\bar{n}$ states to the $b\bar{b}$ mesons within the same model. The quark-quark interaction takes into account QCD perturbative effects by means of the one-gluon exchange potential and the most important nonperturbative effects through the hypothesis of a screened confinement and the spontaneous breaking of chiral symmetry \cite{Vijande_2005}. They have also found evidence of the existence of a $1^{+-}$ light-isoscalar meson with a dominant $s\bar{s}$ content and a mass around 1.97 GeV  and an isoscalar $s\bar{s}$ state at 1.85 GeV \cite{Vijande_2005}.
Anisovich {\it{et al.}}  considered the light-quark ($u, d, s$) mesons with masses $M \leq 3$ GeV; as a result, the calculations have been performed for the mesons lying on linear trajectories in the ($n, M^2$)-planes, where $n$ is the radial quantum number \cite{Anisovich:2005dt}.
In 2023, the mass spectra of the light mesons, the kaons ($u\bar{s}$), and the strangeonium ($s\bar{s}$) were systematically studied within the framework of Regge phenomenology. Several relations between Regge slope, intercept, and meson masses were extracted with the assumption of linear Regge trajectories by Oudichhya {\it{et al.}}  \cite{Oudichhya:2023lva}.
All of the above results are compared with our current work in Table \ref{mass}. We can see that the vast majority of our results are lower than those. A more detailed discussion will be presented in a later section.
 \begin{table*}[htbp]
 \small
 \renewcommand{\arraystretch}{1.5}
\centering
\caption{The mass spectra of $J^{PC}=1^{+-}$ states. The unit of the mass is MeV. \label{mass}}
\[\begin{array}{llcccccccccccc} 
\hline
\hline
&&n^{2s+1}L_J&J^{PC}&\text{State} &\text{This Work}&\text{Expe.}               &\text{GI}$~\cite{Steph:1985ff}$
&$~\cite{PhysRevD.35.265}$&$~\cite{Ebert:2009ub}$ &$~\cite{Li_2021}$   &$~\cite{Vijande_2005}$  &$~\cite{Anisovich:2005dt}$  &$~\cite{Oudichhya:2023lva}$ \\\midrule[1pt]
&         &     & 1^1P_1  &h_1(1170) &1219 &1166\pm6$~\cite{ParticleDataGroup:2024cfk}$   &1218    &1280 &1258   &      &1257  &      &\\
&         &     & 2^1P_1  &h_1(1595) &1699 &1594^{+18}_{-60}$~\cite{BNL-E852:2000poa}$    &1777    &1910 &1721   &      &1700  &      &\\
&         &     & 3^1P_1  &h_1(1965) &2048 &1965\pm45$~\cite{Anisovich:2002xoo}$          &2235    &2390 &2007   &      &      &      &\\
&q\bar{q} &I=0  & 4^1P_1  &h_1(2215) &2320 &2215\pm40$~\cite{Anisovich:2002xoo}$          &2634    &     &2264   &      &      &      &\\
&         &     & 5^1P_1  &h_1(5P)   &2529 &                                            &2986    &     &       &      &      &      &\\
&         &     & 6^1P_1  &h_1(6P)   &2699 &                                            &3312    &     &       &      &      &      & \\
&         &     & 7^1P_1  &h_1(7P)   &2825 &                                            &3607    &     &       &      &      &      &\\\cline{2-14}
&         &     & 1^1P_1  &b_1(1235) &1219 &1229.5\pm3.2$~\cite{ParticleDataGroup:2024cfk}$&1218   &1280 &1258   &      &1234  &1168  &\\
&         &     & 2^1P_1  &b_1(2P)   &1699 &                                             &1777   &1910 &1721   &      &      &1567  &\\
&         &     & 3^1P_1  &b_1(1960) &2048 &1960\pm35~  $\cite{Anisovich:2002xoo}$         &2235   &2390 &2007   &      &      &1928  &\\
&q\bar{q} &I=1  & 4^1P_1  &b_1(2240) &2320 &2240\pm35~  $\cite{Anisovich:2002xoo}$         &2634   &     &2264   &      &      &2240  &\\
&         &     & 5^1P_1  &b_1(5P)   &2529 &                                             &2986   &     &       &      &      &2548  &\\
&         &     & 6^1P_1  &b_1(6P)   &2699 &                                             &3312   &     &       &      &      &2927  &\\
&         &     & 7^1P_1  &b_1(7P)   &2825 &                                             &3607   &     &       &      &      &      &\\\hline
&         &     & 1^1P_1  &h_1(1415) &1479 &1409^{+9}_{-8}$~\cite{ParticleDataGroup:2024cfk}$&1473 &1440 &1485   &1462  &1511  &      &1402.01\pm0.78\\
&         &     & 2^1P_1  &X(1911)   &1951 &1911 \pm 6 \pm 14$~\cite{BESIII:2023zwx}$ ?    &2008   &2030 &2024   &1991  &1973  &      &2024.00\pm0.00\\
&         &     & 3^1P_1  &h_1^\prime(3P) &2298 & 2316 \pm 9\pm30$~\cite{BESIII:2024nhv}$ ?&2449   &2490 &2398   &2435  &      &      &2495.51\pm1.46\\
&s\bar{s} &I=0  & 4^1P_1  &h_1^\prime(4P) &2568 &                                        &2832   &     &2717   &      &      &      &2891.11\pm1.62\\
&         &     & 5^1P_1  &h_1^\prime(5P) &2782 &                                        &3174   &     &       &      &      &      &3238.75\pm1.78\\
&         &     & 6^1P_1  &h_1^\prime(6P) &2956 &                                        &3487   &     &       &      &      &      &\\
&         &     & 7^1P_1  &h_1^\prime(7P) &3096 &                                        &3776   &     &       &      &      &      &\\
 \hline
 \hline
\end{array}\] 
\end{table*}

\subsection{Ground states with \text{$J^{PC}=1^{+-}$}}
There are three states in the family of light $J^{PC}=1^{+-}$, $b_1$, $h_1$ and $h_1^\prime$. 
In 1963, the $b_1(1235)$ was observed as a $\pi\omega$ resonance at around 1.22 GeV in the $\pi^+p \to \pi^+ + p + \pi^+ +\pi^- +\pi^0$ process \cite{Clymton:2023txd}. In 1984, the Omega Photon Collaboration reported masses of $1.222 \pm 0.006$ GeV for the $b_1(1235)^\pm$ and $1.237 \pm 0.007$ for the $b_1(1235)^0$ \cite{OmegaPhoton:1984ols}. In 1991, Fukui {\it{et al.}}  observed a clear $b_1(1235)$ signal in the $1^{+-}$ wave; the fitting of this distribution with the Breit-Wigner form leads to M$=1236 \pm 16$ MeV and $\Gamma = 151 \pm 31$ MeV for the $b_1(1235)$ mass and width, respectively \cite{Fukui:1990ki}. IHEP-IISN-LANL-LAPP-KEK Collaboration obtained $b_1(1235)$ parameters from the 38 GeV and 100 GeV data, M$ = 1235 \pm 15$ MeV, $\Gamma = 160 \pm 30$ MeV in 1992 \cite{IHEP-IISN-LANL-LAPP-KEK:1992puu}. In 1993, ASTERIX Collaboration found $b_1(1235)$ with mass and width are 
$1225 \pm 5$ MeV and $113 \pm 12$ MeV, respectively \cite{ASTERIX:1993wam}.
\par
The $b_1(1235)$ is well established as the ground state in the $b_1$ family \cite{Chen:2015iqa, Clymton:2023txd, Steph:1985ff, Ebert:2009ub, Anisovich:2005dt, Vijande_2005}. The mass of $b_1(1P)$ we calculated is 1219 MeV, and the width of $b_1(1235) \to \omega\pi$ is 147 MeV, which is in agreement with experimental width \cite{ParticleDataGroup:2024cfk}. Moreover, $\omega\pi$ has been seen in the experiment. This experimental observation further substantiates our hypothesis and validates the model we have selected.   

\par
Usually, there are mainly two mixing schemes concerning $h_1-h_1^\prime$ mixing. One is the octet-singlet mixing scheme (denotes as 18), 

\begin{equation}\label{mixing18}
\left( \begin{matrix}
	|h_1\rangle \\
	|h_1^\prime\rangle \\
\end{matrix}\right) =
\left( \begin{matrix}
	\textrm{$\cos\theta_{18}$} & \textrm{$\sin\theta_{18}$}\\
	\textrm{$-\sin\theta_{18}$} & \textrm{$\cos\theta_{18}$} \\
\end{matrix}\right)
\left( \begin{matrix}
	|h_{11}\rangle \\
	|h_{18}\rangle \\
\end{matrix}\right),
\end{equation}     
where $\theta_{18}$ is the vector meson mixing angle in
the octet-singlet mixing scheme. Here, the flavor SU(3) octet basis
is $|h_{18}\rangle = \frac{1}{\sqrt{6}}(u\bar{u}+d\bar{d}-2
s\bar{s})$ and singlet basis is $|h_{11}\rangle =
\frac{1}{\sqrt{3}}(u\bar{u}+d\bar{d}+ s\bar{s})$.
The other is the quark-flavor basis mixing scheme (denoted as $ns$)

\begin{equation}\label{mixingns}
\left( \begin{matrix}
	|h_1\rangle \\
	|h_1^\prime\rangle \\
\end{matrix}\right) =
\left( \begin{matrix}
	\textrm{$\cos\theta_{ns}$} & \textrm{$\sin\theta_{ns}$} \\
	\textrm{$-\sin\theta_{ns}$} & \textrm{$\cos\theta_{ns}$} \\
\end{matrix}\right)
\left( \begin{matrix}
	|n\bar{n}\rangle \\
	|s\bar{s}\rangle \\
\end{matrix}\right),
\end{equation}
where $\theta_{ns}$ is the mixing angle in the quark flavor scheme, and $n\bar{n}=(u\bar{u}+d\bar{d})/\sqrt{2}$.  
The two schemes are equivalent to each other by $\theta_{ns}=\theta_{18}+\text{arctan}(\sqrt{2})$ when the SU(3) symmetry is perfect. 
When SU(3) symmetry is broken, this relationship is no longer maintained  \cite{Qian:2008px}. 

\par 
Here we adopt the quark flavor mixing scheme to study $h_1-{h_1^\prime}$ meson mixing.
For $h_1(1170)$ and $h_1(1415)$, 

\begin{equation}\label{angleh11P}
\left( \begin{matrix}
	|h_1(1170)\rangle \\
	|h_1(1415)\rangle \\
\end{matrix}\right) =
\left( \begin{matrix}
	\textrm{$\cos\theta_{1}$} & \textrm{$\sin\theta_{1}$} \\
	\textrm{$-\sin\theta_{1}$} & \textrm{$\cos\theta_{1}$} \\
\end{matrix}\right)
\left( \begin{matrix}
	|n\bar{n}\rangle \\
	|s\bar{s}\rangle \\
\end{matrix}\right).
\end{equation}

The concrete value of $\theta_1$ is suggested to be $\theta_1\sim82.7^\circ$ \cite{Cheng:2011pb}, $\theta_1=85.6^\circ$ \cite{Li:2005eq}, $\theta_1=86.8^\circ$ \cite{Dudek:2011tt}, $\theta_1=78.7^\circ$ \cite{Wang:2019qyy} by different theoretical groups.
\par
The $h_1(1170)$ was observed by an experiment \cite{Dankowych:1981ks} at ANL in the $\pi^+\pi^-\pi^0$ state of the $\pi^-p$ charge-exchange reaction, it was reported that the mass was $1166 \pm 5 \pm 3$ MeV and the width was $320 \pm 50$ MeV. Ando {\it{et al.}}  have performed a high-statistics experiment on the reaction $\pi^-p \to \pi^+\pi^-\pi^0$ at $8.06$ GeV using a spectrometer detecting both charged particles and gamma rays \cite{Ando:1990ti}. 
As a result, they observed $h_1(1170)$ with mass M$(h_1) = 1166 \pm 5 \pm 3$ MeV and $\Gamma(h_1) = 375 \pm 6 \pm 34$ MeV, respectively  \cite{Ando:1990ti}. Other experimental data can be seen in PDG \cite{ParticleDataGroup:2024cfk}.
According to PDG \cite{ParticleDataGroup:2024cfk}, the $h_1(1170)$ as the $1^1P_1$ isoscalar state (mostly of $u\bar{u} + d\bar{d}$) is well established. 
\par
\begin{figure*}[htbp]
\centering%
\includegraphics[scale=1]{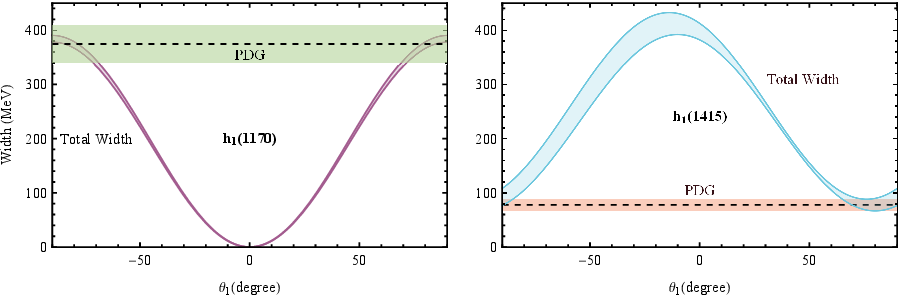}
 \caption{The $\theta_1$ dependence of the total decay widths of $h_1(1170)$ and $h_1(1415)$, where the corresponding experimental data (green and orange band) for comparison with our theoretical calculation are listed.
 The experimental values are from PDG \cite{ParticleDataGroup:2024cfk}. \label{theta1}}
\end{figure*}

The mass of $h_1(1P)$ we obtained is 1219 MeV by the MGI model, which is close to the experimental value \cite{ParticleDataGroup:2024cfk} and lower than other theoretical results as shown in Table \ref{mass}. We give the information on the $\theta_1$ dependence of total decay widths of $h_1(1170)$ as a $1^1P_1$ state in the left-hand image in Fig. \ref{theta1}, where we also compare our result with the experimental data. 
The $h_1(1170)$ mainly decays into $\pi\rho$, which has already been proven by experiments \cite{Ando:1990ti, OmegaPhoton:1983huz, Dankowych:1981ks}.  We take $\theta_1=85.6^\circ$ \cite{Li:2005eq} to calculate the width of $h_1(1170)$ and the result is shown in Table \ref{1pdecay}.
This also verifies the correctness of the model we choose.  
\par
The $h_1(1415)$, which was earlier known as $h_1(1380)$, was a convincing candidate for the $1^1P_1$ $s\bar{s}$ state \cite{Li_2021, Chen:2015iqa, Vijande_2005, Oudichhya:2023lva}.
In 1988, the first evidence of the $^1P_1$ strangeonium state was observed by the LASS spectrometer at SLAC, from a partial wave analysis of $K^-p \to K^0_SK^{\pm}\pi^{\mp}\Lambda$ \cite{Aston:1987ak}, which was confirmed by the crystal barrel detector at LEAR in the $p\bar{p} \to K_LK_S\pi^0\pi^0$ with a mass of $1440 \pm 60$ MeV and a width of $170 \pm 80$ MeV \cite{CrystalBarrel:1997kda}.  
BESIII Collaboration observed a structure ($h_1(1380)$) near the $K^*\bar{K}$ mass threshold in the $K\bar{K}\pi$ invariant mass distribution, and the corresponding mass and width are measured to be M $=1412 \pm 4 \pm 8$ MeV and $\Gamma= 84 \pm 12 \pm 40$ MeV, respectively \cite{BESIII:2015vfb}. 
Subsequently, the BESIII Collaboration measured the mass and width of the $h_1(1415)$ to be M$ = 1423.2 \pm 2.1 \pm 7.1$ MeV and $\Gamma = 90.3 \pm 9.8 \pm 17.5$ MeV, respectively \cite{BESIII:2018ede}. The most precise mass and width of $h_1(1415)$ are measured to be M$ = 1384 \pm 6 ^{+9}_{-8}$ MeV and $\Gamma = 66 + 10^{+12}_{-10}$ MeV by the BESIII Collaboration \cite{BESIII:2022zel}. Due to the severe suppression by the phase space, its dominant decay is $K^*K$. It can also decay into $\pi\rho$ and $\eta\omega$. $h_1(1415)$ mainly decays into $K^*K$, which consistent with the experimental fact that $h_1(1415)$ has a dominant $s\bar{s}$ component \cite{Aston:1987ak, CrystalBarrel:1997kda, BESIII:2015vfb, BESIII:2022zel}. More details can be seen in Table \ref{1pdecay}. The information on the $\theta_1$ dependence of total decay widths of $h_1(1415)$ as a $1^1P_1$ state is in right-hand image in Fig. \ref{theta1}.

\begin{table*}[htbp]
\renewcommand{\arraystretch}{1.5}
\centering
\caption{The total and partial decay widths of the  $1^1P_1$ states, the unit of width is MeV. \label{1pdecay}}
\vspace{-0.2cm}
\[\begin{array}{cccccc}
\hline
\hline
\multicolumn{2}{c}{b_1(1235),~{\Gamma_{exp.}=142 \pm 9} $~\cite{ParticleDataGroup:2024cfk}$}&\multicolumn{2}{c}{h_1(1170),~ {\Gamma_{exp.}=375\pm35} $~\cite{ParticleDataGroup:2024cfk}$}&\multicolumn{2}{c}{h_1(1415),~ {\Gamma_{exp.}=78\pm11} $~\cite{ParticleDataGroup:2024cfk}$}\\
\hline
\text{Channel}&\text{Value}     &\text{Channel}  &\text{Value}  &\text{Channel}  &\text{Value}\\ \midrule[1pt]
\text{Total} & 147 & \text{Total} & 382 & \text{Total} & 85 \\
 \omega \pi  & 147 & \rho \pi  & 382 & K^*K & 82.9 \\
 - & - & - & - & \rho \pi  & 1.75 \\
 \hline
 \hline
\end{array}\]
\end{table*}

\subsection{The first radially excited states with \text{$J^{PC}=1^{+-}$} }
The $b_1(2P)$ is still missing. We make predictions about its mass and decay behavior. From our calculation, the mass of $b_1(2P)$ is about 1700 Mev, which is lower than those in Refs. \cite{Steph:1985ff, PhysRevD.35.265, Ebert:2009ub}. $b_1(2P) \to \pi{a_2}$ will be the dominant decay mode. In our calculation, $\rho\rho~(\to 4\pi)$, $\omega\pi~(\to 4\pi)$, and $\omega(1420)\pi~(\to 4\pi)$ are the important decay channels. The decay modes $\rho \eta$, $\pi a_1$, and $K^*K$ make some contribution to the total width.
We suggest that experimentalists focus on $4\pi$ final channels to find $b_1(2P)$ state. Other decay information can be seen in Table \ref{2pdecay}. 
\par
\begin{figure*}[htbp]
\centering%
\includegraphics[scale=1]{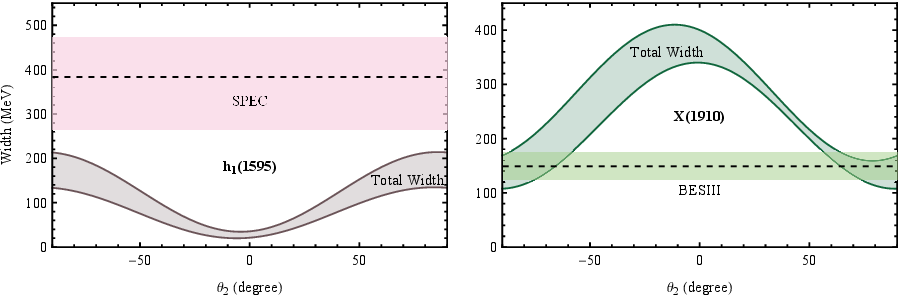}
\caption{The $\theta_2$ dependence of the total decay widths of $h_1(1595)$ and $h_1(1911)$, where the corresponding experimental data (pink and light green band) for comparison with our theoretical calculation are listed. Here, the experimental value are from SPEC \cite{BNL-E852:2000poa} and BESIII \cite{BESIII:2023zwx}. \label{theta2} }
\end{figure*}

According to the Regge trajectory analysis in Fig. \ref{regge} and mass spectra, $h_1(1595)$ is the first radially excited state of $h_1(1170)$. 
$h_1(1595)$ was observed in a particle wave analysis of the $\omega\eta$ final state produced in $\pi^-p$ interactions at 18 GeV/$c$ where $\omega \to \pi^+\pi^-\pi^0$, $\pi^0 \to 2\gamma$, and $\eta \to 2\gamma$ \cite{BNL-E852:2000poa} will correspond to the $n\bar{n}$ $2^1P_1$ state \cite{Vijande_2005, Wang:2019qyy, Chen:2015iqa}.
Here, $h_1(1595)$ has the same flavor wave functions as $h_1(1170)$ in (Eq. \ref{angleh11P}). The mixing angle $\theta_1$ in (Eq. \ref{angleh11P}) is replaced by $\theta_2$ for the corresponding $h_1$ state. As for $h_1(1595)$, the mixing angle $\theta_2$ was not well determined. Thus, we take a typical angle $\theta_2=85.6^{\circ}$ to discuss the decay behavior of $h_1(1595)$. In the case of $h_1(1965)$, $h_1(2215)$, $h_1(5P)$, $h_1(6P)$, and $h_1(7P)$, we employ the same methodology as $h_1(1595)$. The $\rho\pi$ is the dominant decay channel of $h_1(1595)$ as shown in Table \ref{2pdecay}, the width is about 154 MeV, and the branch ratio is 0.88. 
Besides, the $h_1(1595)$ resonance can also decay into $K^*K$ and $\omega\eta$ channels. The $\omega\eta$ channel has been observed in the experiment \cite{BNL-E852:2000poa}. 
The total width of $h_1(1595)$ we obtained is 175 MeV, which is much smaller than experimental width \cite{BNL-E852:2000poa}. We conduct research on the total decay width that $\theta_2$ depends on in the left-hand image in Fig. \ref{theta2} and compare the total width with  the experimental value.
More precise measurements are desperately needed to explain this discrepancy.

\par
The $X(1910)$ was seen in the $J/\psi \to \phi\pi^0\eta$ process, with $J^{PC} = 1^{+-}$, mass M$ = 1911 \pm 6 \pm 14 $ MeV, and width $\Gamma = 149 \pm 12 \pm 23$ MeV at the BEPCII collider by the BESIII Collaboration  \cite{BESIII:2023zwx}. Under the analysis of mass and two-body strong decay behavior, it can be treated as an $s\bar{s}$ $h_1^\prime(2P)$ state. 
For a $2^1P_1$ $s\bar{s}$ state, its mass is 1951 MeV according to our calculation, which is close to the experimental mass of $X(1910)$. 
The total width of $X(1910)$ as a $2^1P_1$ $s\bar{s}$ state we obtained is 133 MeV, which is consistent with the experimental value of $149 \pm 12 \pm 23$ MeV. $K^*K$ is the dominant decay channel as shown in Table \ref{2pdecay}. $K^*K^*$, $K^*(1410)K$, and $\eta\phi$ also have sizable contributions in the total width. 
We present the total width of $X(1910)$ dependent on $\theta_2$ as shown in the right-hand image of Fig. \ref{theta2}. All indications suggest that the primary flavor of the $X(1910)$ is $s\bar{s}$ component.
\par    
The $h_1^\prime(2P)$ has the same flavor wave functions as $h_1(1415)$ in (Eq. \ref{angleh11P}). The mixing angle $\theta_1$ in (Eq. \ref{angleh11P}) is replaced by $\theta_2$ for the corresponding $h_1$ state. As for $h_1^\prime(2P)$, we take a typical angle $\theta_2=85.6^{\circ}$ to discuss its decay behavior. In the case of $h_1^\prime(3P)$, $h_1^\prime(4P)$, $h_1^\prime(5P)$, $h_1^\prime(6P)$, and $h_1^\prime(7P)$, we employ the same methodology as $h_1^\prime(2P)$ state.

\begin{table*}[htbp]
\renewcommand{\arraystretch}{1.5}
\centering
\caption{The total and partial decay widths of the  $2^1P_1$ states, the unit of width is MeV. \label{2pdecay}}
\[\begin{array}{cccccc}
\hline
\hline
\multicolumn{2}{c}{b_1(2P)}&\multicolumn{2}{c}{h_1(1595),~ {\Gamma_{exp.}=384^{+90}_{-120}} ~$\cite{BNL-E852:2000poa}$}&\multicolumn{2}{c}{X(1910),~ {\Gamma_{exp.}=149 \pm 12 \pm 23}$~\cite{BESIII:2023zwx}$ }\\\hline
\text{Channel}   &\text{Value}  &\text{Channel}  &\text{Value}     &\text{Channel}  &\text{Value} \\ \midrule[1pt]
\text{Total} & 395 & \text{Total} & 175 & \text{Total} & 133 \\
 \pi a_2 & 141 & \rho \pi  & 154 & K^*K & 60.8 \\
 \rho \rho  & 81.2 & K^*K & 10.7 & K^*K^* & 39.6 \\
 \omega \pi  & 72 & \omega \eta  & 10.6 & K^*\text{(1410)K} & 21.5 \\
 \text{$\omega $(1420)$\pi $} & 48.2 & - & - & \eta \phi  & 8.32 \\
 \rho \eta  & 22.8 & - & - & - & - \\
 \pi a_1 & 13.2 & - & - & - & - \\
 K^*K & 12.3 & - & - & - & - \\
 \pi a_0 & 4.86 & - & - & - & - \\
\hline
\hline
\end{array}\]
\end{table*}

\subsection{The second radially excited states with \text{$J^{PC}=1^{+-}$} }
The $b_1(1960)$ has been seen in $p\bar{p} \to \omega\pi^0, \omega\eta\pi^0, \pi^+\pi^-$ process according to PDG, whose mass and width are $1960\pm 35$ MeV and $230 \pm 50$ MeV, respectively  \cite{Anisovich:2002su}.
$\pi{a_2}$, $\pi{a_2(1700)}$, $\omega\pi$ are dominant decay channels; the widths are about 35 MeV, 31 MeV, and 30 MeV. $\omega(1420)\pi$ and $\rho\rho$ are important decay channels that have the branch ratio both about 0.1. The final states $\omega(1650)\pi$, $\rho\eta$, and $\pi{a_1}$ also have sizeable decay widths. Other small decay modes can be seen in Table \ref{3pdecay}.  
\par
\begin{figure*}[htbp]
\centering%
\includegraphics[scale=1]{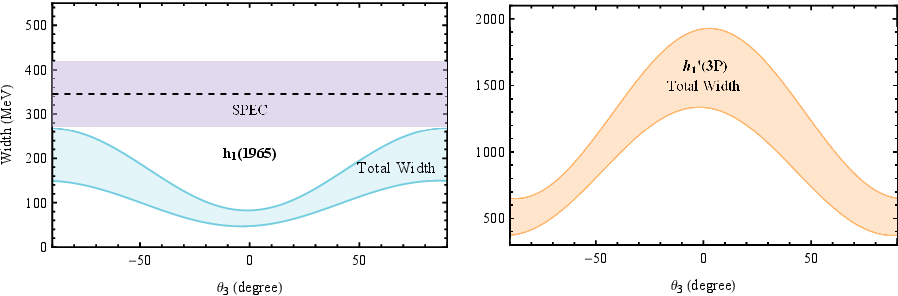}
\caption{The $\theta_3$ dependence of the total decay widths of $h_1(1965)$ and $X(2300)$, where the corresponding experimental data of $h_1(1965)$ (purple band) for comparison with our theoretical calculation is listed. Here, the experimental value is from SPEC \cite{Anisovich:2002xoo}. \label{theta3}}
\end{figure*}

The $h_1(1965)$ is from a combined fit to $\omega\eta$ and $\omega\pi^0\pi^0$, using both $\omega \to \pi^0\gamma$ and $\omega \to \pi^+\pi^-\pi^0$ decays \cite{Anisovich:2002xoo}. We carried out the total width depending on $\theta_3$ as shown in the left-hand image in Fig. \ref{theta3}. $h_1(1965)$ has the same flavor wave functions as $h_1(1170)$ in (Eq. \ref{angleh11P}) as mentioned. We substitute the mixing angle 
$\theta_1$ in (Eq. \ref{angleh11P}) for $\theta_3$. We present our result when taking $\theta_3=85.6^\circ$ in Table \ref{3pdecay}. $h_1(1965)$ dominantly decays into $\rho\pi$ if we regard it as a $n\bar{n}$ $3^1P_1$ state. The decay modes $\rho(1450)\pi$, $\rho(1700)\pi$, $K^*(1410)K$, and $\pi{\rho_3(1690)}$ make some contribution to the total width. According to our calculation, the total width of $h_1(1965)$ is 202 MeV, which is a little lower than the experimental data  \cite{Anisovich:2002xoo}.     
\par
The BESIII Collaboration observed the $X(2300)$, determining its mass and width to be $2316 \pm 9 \pm 30$ MeV and $89 \pm 15 \pm 26$ MeV, respectively \cite{BESIII:2024nhv}.
The total width of the $X(2300)$, which depends on $\theta_3$, is depicted in the right-hand image of Fig. \ref{theta3}.
A comparison between the predicted results and the available experimental data for the $X(2100)$ and the $X(2300)$ indicates that, without further confirmation of these states, it may be premature to assign either the $X(2100)$ or the$X(2300)$ as the $s\bar{s}$ member of the $^1P_1$ meson.
Based on Regge trajectory and the result of the MGI model, it is reasonable to classify the $X(2300)$ as a $s\bar{s}$ $3^1P_1$ state within the mass spectrum.
However, for the $s\bar{s}$ $3^1P_1$ state, the theoretical total width calculated using the $^3P_0$ model significantly exceeds that of the $X(2300)$. Further theoretical and experimental research is required to elucidate its structure.
In contrast, the $X(2100)$ has a measured mass of M$ = 2062.8 \pm 13.1 \pm 4.2$ MeV \cite{BESIII:2018zbm}, which is significantly below the MGI model's prediction of 2297.6 MeV. In terms of total width, the $X(2100)$ has a total width of $\Gamma = 177 \pm 36 \pm 20$ MeV. In contrast, our QPC model result for the $s\bar{s}$ $h_1^\prime(3P)$ is 284 MeV. As a result, the structure of $h_1^\prime(3P)$ remains unclear.

\begin{table*}[htbp]
\renewcommand{\arraystretch}{1.5}
\centering
\caption{The total and partial decay widths of the  $3^1P_1$ states, the unit of width is MeV. \label{3pdecay}}
\[\begin{array}{cccccc}
\hline
\hline
\multicolumn{2}{c}{b_1(1960),~ {\Gamma_{exp.}=230\pm50} $~\cite{Anisovich:2002xoo}$}&\multicolumn{2}{c}{h_1(1965),~ {\Gamma_{exp.}=345\pm75}$~\cite{Anisovich:2002xoo}$ }&\multicolumn{2}{c}{h_1^\prime(3P) }\\\hline
\text{Channel}             &\text{Value}    &\text{Channel}  &\text{Value}     &\text{Channel}  &\text{Value} \\ \midrule[1pt]
\text{Total} & 192          & \text{Total} & 202      & \text{Total} & 284 \\
 \pi a_2 & 34.9               & \rho \pi  & 93.3 & K^*\text{(1410)}K^* & 71.3 \\
 \pi a_2\text{(1700)} & 31.2 & \text{$\rho $(1450)$\pi $} & 51.3 & K^*K & 54.4 \\
 \omega \pi  & 29.6 & \text{$\rho $(1700)$\pi $} & 21.8 & KK_2{}^* & 49.1 \\
 \text{$\omega $(1420)$\pi $} & 20.5  & K^*\text{(1410)K} & 8.45 & K^*K^* & 29.5 \\
 \rho \rho  & 19.8 & \pi \rho _3\text{(1690)} & 7.28 & K_1K^* & 21 \\
 \text{$\omega $(1650)$\pi $} & 8.54 & \omega \eta  & 6.74 & K^*\text{(1410)K} & 15.6 \\
 \rho \eta  & 6.79       & K^*K & 6.42           & K^*\text{(1680)K} & 10.6 \\
 \pi a_1 & 6.67          & K^*K^* & 4.84 & {K_1^\prime}K^* & 6.71 \\
 K^*\text{(1410)K} & 6.51 & \text{$\omega \eta^\prime$} & 1.65 & \eta \phi  & 6.31 \\
 \pi a_0\text{(1740)} & 6.42 & - & - & K_3{}^*\text{(1780)K} & 4.34 \\
 K^*K^* & 6.1 & - & - & \text{$\eta \phi $(1680)} & 3.05 \\
 K^*K & 5.48 & - & - & \text{$\eta $(1295)$\phi $} & 2.15 \\
 \pi a_1\text{(1640)} & 3.9 & - & - & \text{$\eta^\prime  \phi $} & 2.12 \\
 \pi \omega _3\text{(1670)} & 3.48 & - & - & {K_1^\prime}K & 1.3 \\
 \text{$\rho \eta^\prime$} & 1.66 & - & - & \rho \pi  & 1.08 \\
 \hline
 \hline
\end{array}\]
\end{table*}

\subsection{The third radially excited states with \text{$J^{PC}=1^{+-}$} }
As for the third  radially excited states of the $J^{PC}=1^{+-}$ family, we present their total and partial decay widths in Table \ref{4pdecay}. Assuming that $b_1(2240)$ is the third radial radially excited state, its experiment width of $320 \pm 85$ MeV \cite{Anisovich:2002xoo} can cover our calculated result of 252 MeV. 
$\pi{a_2(1950)}$ will be the main decay channel of $b_1(2240)$, and the branch ratio is about 0.3. $\pi{a_2(1700)}$, $\omega(1420)\pi$, $\pi{a_2}$, and $\rho{b_1}$ are important decay channels for $b_1(2240)$ as a $4^1P_1$ state.
$\omega\pi$, $\pi\omega_3(1945)$, $\omega{a_1}$, $\rho\rho$, $\pi\omega(2D)$, and $\rho{f_2}$ also have sizable decay widths.
\par
\begin{figure}[htbp]
\centering%
\includegraphics[scale=0.75]{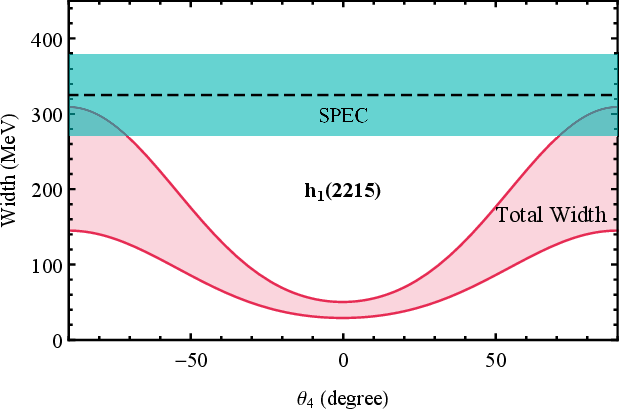}
\caption{The $\theta_4$ dependence of the total decay widths of $h_1(2215)$, where the corresponding experimental data (turquoise blue band) for comparison with our theoretical calculation is listed.
Here, the experimental value is from SPEC \cite{Anisovich:2002xoo}. }
\label{2215}
\end{figure}

The $h_1(2215)$, with the mass of $2215 \pm 40$ MeV and the width of $325 \pm 55$ MeV \cite{Anisovich:2002xoo}, is commonly considered to be in the $4^1P_1$ state. $\rho(1450)\pi$, $\rho\pi$, $\rho{a_1}$, and $\rho(2000)\pi$ are dominant decay channels with branch ratios of 0.19, 0.13, 0.12, and 0.11, respectively. Figure \ref{2215} illustrates the $\theta_4$ dependence of the total decay width of the $h_1(2215)$, demonstrating that our theoretical predictions are in good agreement with the experimental data.

The $h_1^\prime(4P)$ is the $s\bar{s}$ partner of $h_1(2215)$, has the mass of 2568 MeV and the width of 260 MeV in our prediction. $h_1^\prime(4P)$ mainly decays to two strange mesons for its $s\bar{s}$ component. $K^*K$ and $KK_2^*$ are the dominant decay modes with the widths 40 MeV and 37 MeV, respectively. $K^*(1410)K$ and $K^*K_1(1650)$ are also  important decay channels with the branch ratios 0.1 and 0.05. Besides, $K^*K(1630)$, $K^*K^*$, $K^*(1410)K$, $KK^*(3S)$, $K_3^*(1780)K$, and $K_1K^*$ are its sizable final channels with width over 10 MeV. $K^*K_2^*$, $KK_2^*(1980)$, ${K_1^\prime}K^*$, $K^*K(1460)$ are the visible decay modes of $h_1^\prime(4P)$ as shown in Table \ref{4pdecay}.

\begin{table*}[htbp]
\renewcommand{\arraystretch}{1.4}
\centering
\caption{The total and partial decay widths of the  $4^1P_1$ states, the unit of width is MeV. \label{4pdecay}}
\[\begin{array}{cccccc}
\hline
\hline
\multicolumn{2}{c}{b_1(2240),~ {\Gamma_{exp.}=320\pm85} $~\cite{Anisovich:2002xoo}$}&\multicolumn{2}{c}{h_1(2215),~ {\Gamma_{exp.}= 325\pm55}$~\cite{Anisovich:2002xoo}$ }&\multicolumn{2}{c}{h_1^\prime(4P) }\\\hline
\text{Channel}                &\text{Value}   &\text{Channel}     &\text{Value}    &\text{Channel}       &\text{Value}  \\\midrule[1pt]
\text{Total} & 252 & \text{Total} & 178 & \text{Total} & 259 \\
 \pi a_2\text{(1950)} & 74.6 & \text{$\rho $(1450)$\pi $} & 34.7 & K^*K & 39.7 \\
 \pi a_2\text{(1700)} & 29.9 & \rho \pi  & 23.1 & KK_2{}^* & 37.2 \\
 \text{$\omega $(1420)$\pi $} & 18.5 & \rho a_1 & 21.2 & K^*\text{(1410)}K^* & 27 \\
 \pi a_2 & 14.9 & \text{$\rho $(2000)$\pi $} & 20.4 & K^*K_1\text{(1650)} & 24 \\
 \rho b_1 & 12.2 & \text{$\rho $(1900)$\pi $} & 13.2 & K^*\text{K(1630)} & 16 \\
 \omega \pi  & 9.1 & \text{$\rho $(1700)$\pi $} & 12.5 & K^*K^* & 15.9 \\
 \pi \omega _3\text{(1945)} & 8.53 & \text{$\rho \pi $(1300)} & 10.7 & K^*\text{(1410)K} & 14.3 \\
 \omega a_1 & 8.09 & \rho a_2 & 7.87 & KK^*\text{(3S)} & 12.2 \\
 \rho \rho  & 7.92 & \pi \rho _3\text{(1990)} & 6.2 & K_3{}^*\text{(1780)K} & 11.7 \\
 \text{$\pi \omega $(2D)} & 7.5 & K^*\text{(1410)K} & 5.56 & K_1K^* & 11 \\
 \rho f_2 & 7.39 & \omega f_2 & 5.06 & K^*K_2{}^* & 7.75 \\
 \rho f_1 & 5.44 & \omega f_1 & 3.66 & KK_2{}^*\text{(1980)} & 7.35 \\
 K^*\text{(1410)K} & 5.11 & \pi \rho _3\text{(1690)} & 3.63 & {K_1^\prime}K^* & 5.98 \\
 \pi a_1\text{(1640)} & 5.1 & \text{$\omega \eta $(1295)} & 2.15 & K^*\text{K(1460)} & 5.21 \\
 \text{$\omega $(1650)$\pi $} & 4.48 & \text{$\omega $(1420)$\eta $} & 2.08 & K^*\text{(1680)K} & 3.94 \\
 \omega a_2 & 3.91 & K^*K & 1.82 & \eta \phi  & 2.3 \\
 \pi a_0\text{(3P)} & 3.71 & \omega \eta  & 1.53 & \text{$\eta \phi $(1D)} & 2.17 \\
 \text{$\omega \pi $(1300)} & 3.68 & K^*K^* & 1.13 & f_1\text{(1420)$\phi $} & 2.05 \\
 \text{$\omega $(1960)$\pi $} & 3.36 & - & - & \text{$\eta $(1295)$\phi $} & 1.71 \\
 \pi \omega _3\text{(1670)} & 2.74 & - & - & \text{$\eta \phi $(1680)} & 1.7 \\
 \pi a_1 & 2.71 & - & - & \text{$\eta $(1475)$\phi $} & 1.51 \\
 \text{$\rho \eta $(1295)} & 2.22 & - & - & - & - \\
 \text{$\rho $(1450)$\eta $} & 2.07 & - & - & - & - \\
 \rho \eta  & 1.94 & - & - & - & - \\
 K^*\text{(1680)K} & 1.87 & - & - & - & - \\
 K^*K^* & 1.81 & - & - & - & - \\
 K^*K & 1.67 & - & - & - & - \\
 \pi a_0 & 1.31 & - & - & - & - \\
 \hline
  \hline
\end{array}\]
\end{table*}

\subsection{The fourth, fifth and sixth radially excited states with \text{$J^{PC}=1^{+-}$}  }
In this section, we will analyze the fourth, fifth, and sixth  radially excited states of the $J^{PC}=1^{+-}$ family. 
\par
As $5^1P_1$ states in the $J^{PC}=1^{+-}$ family, $b_1(5P)$, $h_1(5P)$, and $h_1^\prime(5P)$ have not yet been observed in any experimental studies. We will make a prediction of their masses and OZI-allowed two-body strong decay behaviors.
\par
Our calculations reveal that both $b_1(5P)$ and $h_1(5P)$  possess identical masses, each being 2530 MeV. The mass of $h_1^\prime(5P)$ is 2780 MeV. 
The decay information of the $b_1(5P)$, $h_1(5P)$, and $h_1^\prime(5P)$ are shown in Table \ref{5pdecay}. The total width of $b_1(5P)$ is approximately 390 MeV. The channels $\pi{a_2}(2175)$, $\pi{a_2}(1950)$, and $\pi{a_2}(1700)$ have the branching ratios of 0.2, 0.14, and 0.12, respectively, which are the main decay modes. $\pi\omega_3(1945)$, $\rho(1450)\rho$, $\omega(1420)\pi$, and $\pi{a_0(4P)}$ are its important decay channels. Otherwise, $\pi{a_2}$, $\rho{b_1}$, $\pi{\omega_3(2285)}$, $\rho{f_2}$, and $\omega{a_2}$ have sizable contributions to the total width of $b_1(5P)$. 
\par
The $h_1(5P)$ and the $h_1^\prime(5P)$ will be $n\bar{n}$ and $s\bar{s}$ partners. 
The total width of $h_1(5P)$ is 340 MeV. The most important decay channels are $\rho(1450)\pi$, $\pi{\rho_3(1990)}$, $\pi\rho_3(2250)$, and $\rho(1900)\pi$ for $h_1(5P)$. $\rho{a_2}$, $\rho{a_1}$, $\rho{a_1(1640)}$, $\pi\rho(4S)$, $\pi\rho(3D)$, $\rho\pi$, and $\rho(2000)\pi$ are also the sizable decay modes. Other decay channels like $\rho\pi(1300)$, $\pi\rho_3(1690)$, $\omega{f_2}$, $\rho(1700)\pi$, and $\omega{f_1}$ are small. More information can be seen in Table \ref{5pdecay}.
As for $h_1^\prime(5P)$, the total width is 290 MeV, in which the decay modes $KK_2^*$, $K^*K$, $K^*(1410)K^*$, and $K^*K(1830)$ make a great contribution. $K^*K(1630)$, $K^*K_1^\prime(2P)$, $K_3^*(1780)K$, and $K^*(1410)K$ are important decay channels for $h_1^\prime(5P)$.

\begin{table*}[htbp]
\renewcommand{\arraystretch}{1.1}
\centering
\caption{The total and partial decay widths of the  $5^1P_1$ states, the unit of width is MeV. \label{5pdecay}}
\[\begin{array}{cccccc}
\hline
\hline
\multicolumn{2}{c}{b_1(5P)}&\multicolumn{2}{c}{h_1(5P)}&\multicolumn{2}{c}{h_1^\prime(5P) }\\\hline
\text{Channel}                &\text{Value}   &\text{Channel}              &\text{Value}      &\text{Channel}   &\text{Value}  \\\midrule[1pt]
\text{Total} & 393 & \text{Total} & 337 & \text{Total} & 288 \\
 \pi a_2\text{(2175)} & 76.7 & \text{$\rho $(1450)$\pi $} & \color{black}{38.8} & KK_2{}^* & 29.1 \\
 \pi a_2\text{(1950)} & 55.7 & \pi \rho _3\text{(1990)} & \color{black} {38.6} & K^*K & 28.5 \\
 \pi a_2\text{(1700)} & 47.3 & \pi \rho _3\text{(2250)} & \color{black}{35.3} & K^*\text{(1410)}K^* & 27.2 \\
 \pi \omega _3\text{(1945)} & 18.7 & \text{$\rho $(1900)$\pi $} & \color{black}{34.6} & K^*\text{K(1830)} & 23.9 \\
 \text{$\rho $(1450)$\rho $} & 18.2 & \rho a_2 & 23.7 & K^*\text{K(1630)} & 18.1 \\
 \text{$\omega $(1420)$\pi $} & 16.8 & \rho a_1 & 20 & K^*K_1'\text{(2P)} & 15.1 \\
 \pi a_0\text{(4P)} & 10.3 & \rho a_1\text{(1640)} & 19.9 & K_3{}^*\text{(1780)K} & 13.9 \\
 \pi a_2 & 9.65 & \text{$\pi \rho $(4S)} & 19.5 & K^*\text{(1410)K} & 13.5 \\
 \rho b_1 & 9.32 & \text{$\pi \rho $(3D)} & 17 & KK_2{}^*\text{(1980)} & 12.9 \\
 \pi \omega _3\text{(2285)} & 8.5 & \rho \pi  & 14.2 & KK^*\text{(3S)} & 12.4 \\
 \rho f_2 & 8.18 & \text{$\rho $(2000)$\pi $} & 10.8 & K_1K^*\text{(1410)} & 10.7 \\
 \omega a_2 & 7.83 & \text{$\rho \pi $(1300)} & 8.95 & K^*K^* & 10.1 \\
 \text{$\omega $(1960)$\pi $} & 7.1 & \pi \rho _3\text{(1690)} & 8.76 & K^*\text{K(1460)} & 8.45 \\
 \omega a_1 & 6.73 & \omega f_2 & 8.14 & K_1K^* & 7.12 \\
 \text{$\pi \omega $(4S)} & 6.53 & \text{$\rho $(1700)$\pi $} & 7.73 & K^*K_1\text{(1650)} & 5.33 \\
 \rho f_1 & 6.15 & \omega f_1 & 6.07 & KK_2{}^*\text{(3P)} & 5.29 \\
 \omega a_1\text{(1640)} & 6.04 & KK^*\text{(3S)} & 3.13 & {K_1^\prime}K^* & 4.55 \\
 \text{$\pi \omega $(3D)} & 5.69 & K^*\text{(1410)}K^* & 2.74 & KK_3{}^*\text{(2D)} & 4.55 \\
 \pi a_1\text{(1640)} & 4.87 & K^*\text{(1680)K} & 2.62 & KK_1{}^*\text{(2D)} & 4.48 \\
 \rho \rho  & 4.7 & K^*\text{(1410)K} & 2.43 & K^*K_2{}^* & 3.88 \\
 \omega \pi  & 4.6 & \text{$\omega $(1420)$\eta $} & 2.4 & K_4\text{(2045)K} & 2.53 \\
 \pi a_0\text{(3P)} & 4.45 & \text{$\omega \eta $(1295)} & 1.76 & f_1\text{(1420)$\phi $} & 2.41 \\
 \pi \omega _3\text{(1670)} & 3.92 & K^*\text{K(1630)} & 1.62 & K^*\text{(1680)K} & 2.16 \\
 K^*\text{(1410)}K^* & 3.91 & \text{$\omega $(1650)$\eta $} & 1.59 & K_3{}^*\text{(1780)}K^* & 1.68 \\
 \text{$\pi \omega $(2D)} & 3.51 & K^*\text{K(1460)} & 1.47 & K^*K_2\text{(1770)} & 1.57 \\
 \rho b_1\text{(2P)} & 3.12 & - & - & KK_2{}^*\text{(1F)} & 1.53 \\
 \text{$\omega \pi $(1300)} & 2.79 & - & - & K^*\text{(1680)}K^* & 1.37 \\
 KK^*\text{(3S)} & 2.66 & - & - & \text{$\eta \phi $(1D)} & 1.26 \\
 \pi a_1\text{(3P)} & 2.6 & - & - & \text{$\eta \phi $(3S)} & 1.02 \\
 \text{$\omega $(1650)$\pi $} & 2.43 & - & - & - & - \\
 K^*\text{(1410)K} & 2.39 & - & - & - & - \\
 \pi a_4\text{(2255)} & 2.26 & - & - & - & - \\
 \pi a_2\text{(2F)} & 2.13 & - & - & - & - \\
 K^*\text{(1680)K} & 2.05 & - & - & - & - \\
 \text{$\rho \eta $(1295)} & 1.9 & - & - & - & - \\
 \pi a_1 & 1.8 & - & - & - & - \\
 \text{$\rho $(1900)$\eta $} & 1.72 & - & - & - & - \\
 \text{$\rho $(1450)$\eta $} & 1.61 & - & - & - & - \\
 K^*\text{K(1630)} & 1.6 & - & - & - & - \\
 \text{$\rho $(1700)$\eta $} & 1.32 & - & - & - & - \\
 \pi a_2\text{(2030)} & 1.17 & - & - & - & - \\
 K^*\text{K(1460)} & 1.13 & - & - & - & - \\
 \rho \rho _2\text{(1D)} & 1.07 & - & - & - & - \\
 \pi a_4\text{(2040)} & 1.07 & - & - & - & - \\
 \rho \eta  & 1.02 & - & - & - & - \\
 \hline
  \hline
\end{array}\]
\end{table*}

As shown in Fig. \ref{regge} and Table \ref{6pdecay}, the masses and decay widths of $b_1(6P)$, $h_1(6P)$, and $h_1^\prime(6P)$ are predicted, which are still unobserved. 
$b_1(6P)$ has the mass 2700 MeV and the total width of about 280 MeV. $\pi{a_2(1950)}$ is its dominant decay channel, whose branch ratio is 0.16. $\pi{a_2(2175)}$, $\pi{a_2(1700)}$, $\pi{\omega_3(1945)}$, $\rho(1450)\rho$ are its important decay channels, and more details can be seen in Table \ref{6pdecay}.
\par
The $h_1(6P)$ and the $h_1^\prime(6P)$ still have the mixing scheme like Eq. \ref{mixingns}. The masses of $h_1(6P)$ and $h_1^\prime(6P)$ will be 2700 MeV and 2956 MeV, respectively. Meanwhile, we make predictions that the width of $h_1(6P)$ is about 280 MeV, and $h_1^\prime(6P)$ is approximately 265 MeV. The main decay modes of $h_1(6P)$ are $\pi{\rho_3(2250)}$, $\pi{\rho_3(1990)}$, $\rho(1900)\pi$, $\pi\rho(4S)$, and $\rho(1450)\pi$. $h_1^\prime(6P)$ mainly decay into $K^*(1410)K$, $KK_2^*$, and $K^*K$. $KK_2^*(1980)$, $K_3^*(1780)K$, $KK_2^*(3P)$, $K^*(1410)K$, and $KK_3^*(2D)$ are its important decay channels. These predictions can help us search for $b_1(6P)$, $h_1(6P)$, and $h_1^\prime(6P)$.         

\begin{table*}[htbp]
\renewcommand{\arraystretch}{1.1}
\vspace{-0.8cm}
\caption{The total and partial decay widths of the  $6^1P_1$ states, the unit of width is MeV. \label{6pdecay}}
\[\begin{array}{cccccc}
\hline
\hline
\multicolumn{2}{c}{b_1(6P)}&\multicolumn{2}{c}{h_1(6P)}&\multicolumn{2}{c}{h_1^\prime(6P) }\\\hline
\text{Channel}                &\text{Value}   &\text{Channel}              &\text{Value}      &\text{Channel}   &\text{Value}  \\\midrule[1pt]
\text{Total} & 279 & \text{Total} & 282 & \text{Total} & 265 \\
 \pi a_2\text{(1950)} & 46 & \pi \rho _3\text{(2250)} & 36.2 & K^*\text{(1410)}K^* & 24.5 \\
 \pi a_2\text{(2175)} & 34.7 & \pi \rho _3\text{(1990)} & 28.6 & KK_2{}^* & 22.9 \\
 \pi a_2\text{(1700)} & 29.8 & \text{$\rho $(1900)$\pi $} & 24.7 & K^*K & 20.6 \\
 \pi \omega _3\text{(1945)} & 14 & \text{$\pi \rho $(4S)} & 21.4 & KK_2{}^*\text{(1980)} & 13.9 \\
 \text{$\rho $(1450)$\rho $} & 10.2 & \text{$\rho $(1450)$\pi $} & 20.9 & K_3{}^*\text{(1780)K} & 13.1 \\
 \pi \omega _3\text{(2285)} & 9.44 & \rho a_2 & 15.6 & KK_2{}^*\text{(3P)} & 12.6 \\
 \text{$\omega $(1420)$\pi $} & 9.39 & \rho a_1\text{(1640)} & 15.4 & K^*\text{(1410)K} & 11.7 \\
 \text{$\pi \omega $(4S)} & 7.19 & \rho a_1 & 12.8 & KK_3{}^*\text{(2D)} & 10.2 \\
 \rho b_1\text{(2P)} & 7.06 & \rho a_2\text{(1700)} & 11.1 & KK^*\text{(3S)} & 9.94 \\
 \rho b_1 & 5.87 & \text{$\pi \rho $(4D)} & 10.8 & K^*\text{K(1630)} & 9.27 \\
 \rho f_2 & 5.43 & \text{$\pi \rho $(5S)} & 8.71 & K^*\text{K(1460)} & 8.66 \\
 \omega a_2 & 5.17 & \text{$\pi \rho $(3D)} & 6.38 & K_1K^*\text{(1410)} & 8 \\
 \omega a_1\text{(1640)} & 5.14 & \rho \pi  & 5.49 & K^*K^* & 7.14 \\
 \pi a_0\text{(5P)} & 4.94 & \omega f_2 & 5.4 & K^*K_1'\text{(2P)} & 6.35 \\
 \text{$\omega $(1960)$\pi $} & 4.7 & \text{$\rho \pi $(1300)} & 4.78 & K_1K_2{}^* & 6.3 \\
 \pi a_0\text{(4P)} & 4.38 & \text{$\rho $(2000)$\pi $} & 4.43 & K^*\text{(1410)}K^*\text{(1410)} & 5.1 \\
 \omega a_1 & 4.32 & \text{$\rho $(1700)$\pi $} & 4.29 & KK^*\text{(4S)} & 4.99 \\
 \rho f_1 & 3.99 & \pi \rho _3\text{(4D)} & 4.08 & K_1K^* & 4.76 \\
 \rho f_1\text{(2P)} & 3.75 & \text{$\rho \pi $(1800)} & 3.97 & KK_1{}^*\text{(3D)} & 4.7 \\
 \text{$\pi \omega $(4D)} & 3.61 & \omega f_1 & 3.94 & K^*\text{K(1830)} & 4.36 \\
  \omega a_2\text{(1700)} & 3.43 & \pi \rho _3\text{(1690)} & 3.59 & {K_1^\prime}K^* & 4.17 \\
 \pi a_4\text{(2255)} & 3.42 & \omega f_1\text{(2P)} & 3.51 & K_4\text{(2045)K} & 4.03 \\
 \rho f_2\text{(2P)} & 3.39 & \omega f_2\text{(2P)} & 3.07 & K^*K^*\text{(3S)} & 3.8 \\
 \text{$\pi \omega $(5S)} & 2.92 & \rho \pi _2 & 3.02 & K^*K_2\text{(1820)} & 2.91 \\
 \rho \rho _2\text{(1D)} & 2.85 & h_1f_2 & 2.29 & K^*K_2{}^* & 2.88 \\
 \rho \rho  & 2.85 & K^*\text{K(1630)} & 1.68 & K_3{}^*\text{(1780)}K^* & 2.85 \\
 \pi a_2 & 2.66 & KK^*\text{(3S)} & 1.61 & {K_1^\prime}K^*\text{(1410)} & 2.76 \\
 \pi a_1\text{(1640)} & 2.62 & KK_1{}^*\text{(2D)} & 1.34 & K_1K_0\text{(1430)} & 2.32 \\
 \pi a_0\text{(3P)} & 2.43 & K^*\text{(1680)K} & 1.31 & K^*\text{(1410)K(1460)} & 2.13 \\
 \text{$\pi \omega $(3D)} & 2.14 & \text{$\eta \omega $(2D)} & 1.31 & K^*K_2\text{(1770)} & 2.12 \\
 a_1a_1 & 1.91 & \text{$\omega $(1420)$\eta $} & 1.24 & K^*\text{(1410)}K_2{}^* & 1.99 \\
 \pi a_1\text{(3P)} & 1.79 & \omega \eta _2 & 1.13 & KK_1{}^*\text{(2D)} & 1.76 \\
 \omega \pi  & 1.77 & K^*\text{(1410)}K^* & 1.13 & {K_1^\prime}K_1 & 1.73 \\
 \pi \omega _3\text{(1670)} & 1.66 & - & - & f_1\text{(1420)$\phi $} & 1.72 \\
 K^*\text{(1410)}K^* & 1.65 & - & - & KK_2{}^*\text{(2F)} & 1.7 \\
 \rho \rho _3\text{(1690)} & 1.58 & - & - & KK_0{}^*\text{(4P)} & 1.64 \\
 \pi a_2\text{(2F)} & 1.54 & - & - & K^*K_2{}^*\text{(1980)} & 1.57 \\
 \pi a_2\text{(5P)} & 1.54 & - & - & K^*\text{(1680)}K^* & 1.51 \\
 \pi a_2\text{(3F)} & 1.52 & - & - & K^*K_1\text{(1650)} & 1.48 \\
 \text{$\omega \pi $(1300)} & 1.48 & - & - & KK_3{}^*\text{(3D)} & 1.23 \\
 \text{$\pi \omega $(2D)} & 1.42 & - & - & K^*\text{(1680)K} & 1.21 \\
 K^*\text{K(1630)} & 1.41 & - & - & KK_2{}^*\text{(1F)} & 1.04 \\
 \pi \omega _3\text{(4D)} & 1.37 & - & - & - & - \\
 KK^*\text{(3S)} & 1.31 & - & - & - & - \\
 \text{$\omega $(1650)$\pi $} & 1.29 & - & - & - & - \\
 \text{$\omega \pi $(1800)} & 1.29 & - & - & - & - \\
 \text{$\rho $(2000)$\eta $} & 1.23 & - & - & - & - \\
 \rho \eta _2 & 1.21 & - & - & - & - \\
 \pi a_1\text{(4P)} & 1.16 & - & - & - & - \\
 KK_1{}^*\text{(2D)} & 1.15 & - & - & - & - \\
 h_1a_2 & 1.09 & - & - & - & - \\
 \text{$\rho \eta $(1295)} & 1.02 & - & - & - & - \\
 \hline
  \hline
\end{array}\]
\end{table*}

Subsequently, we will predict some information about $b_1(7P)$, $h_1(7P)$, and $h_1^\prime(7P)$. Their masses will be 2825 MeV, 2825 MeV, and 3096 MeV by the MGI model. $\pi{a_2(1950)}$ is predicted to be the important decay channel of $b_1(7P)$, which has  ratios of 0.18. Besides, $b_1(7P)$ dominantly decays into $\pi{a_2(1700)}$, $\pi{a_2(2175)}$, $\pi{\omega_3(1945)}$ and so on.
The partial width and total width of $h_1(7P)$, and $h_1^\prime(7P)$ are listed in Table \ref{7pdecay}. The total decay width of $h_1(7P)$ is expected to be $\Gamma_{h_1(7^1P_1)} = 222$ MeV, and the main decay modes are predicted to be $\pi{\rho_3(2250)}$, $\pi{\rho_3(1990)}$, $\pi{\rho(4S)}$, $\rho(1900)\pi$, $\rho(1450)\pi$, and $\pi\rho(5S)$. The total width of $h_1^\prime(7P)$ we predicted will be 231 MeV. $h_1^\prime(7P)$ mainly decays into $K^*(1410)K^*$, $KK_2^*$, $K^*K$, $KK_2^*(3P)$ and $KK_2^*(1980)$.

\begin{table*}[htbp]
\renewcommand{\arraystretch}{1.1}
\vspace{-0.8cm}
\caption{The total and partial decay widths of the  $7^1P_1$ states, the unit of width 
is MeV. \label{7pdecay}}
\[\begin{array}{cccccc}
\hline
\hline
\multicolumn{2}{c}{b_1(7P)}   &\multicolumn{2}{c}{h_1(7P)}                               &\multicolumn{2}{c}{h_1^\prime(7P) }\\\hline
\text{Channel}                &\text{Value}   &\text{Channel}              &\text{Value}      &\text{Channel}   &\text{Value}  \\\midrule[1pt]
\text{Total} & 204 & \text{Total} & 222 & \text{Total} & 231 \\
 \pi a_2\text{(1950)} & 36.7 & \pi \rho _3\text{(2250)} & 28.8 & K^*\text{(1410)}K^* & 21.1 \\
 \pi a_2\text{(1700)} & 19.2 & \pi \rho _3\text{(1990)} & 20.9 & KK_2{}^* & 17.9 \\
 \pi a_2\text{(2175)} & 17.9 & \text{$\pi \rho $(4S)} & 18.2 & K^*K & 15.1 \\
 \pi \omega _3\text{(1945)} & 10.2 & \text{$\rho $(1900)$\pi $} & 17.5 & KK_2{}^*\text{(3P)} & 13 \\
 \pi a_2\text{(5P)} & 8.19 & \text{$\rho $(1450)$\pi $} & 12.8 & KK_2{}^*\text{(1980)} & 12.9 \\
 \pi \omega _3\text{(2285)} & 7.8 & \text{$\pi \rho $(5S)} & 10.4 & KK_3{}^*\text{(2D)} & 11.5 \\
 \text{$\pi \omega $(4S)} & 6.1 & \pi \rho _3\text{(4D)} & 9.38 & K_3{}^*\text{(1780)K} & 11.3 \\
 \text{$\omega $(1420)$\pi $} & 5.79 & \rho a_2\text{(1700)} & 9.3 & K^*\text{(1410)K} & 9.26 \\
 \text{$\rho $(1450)$\rho $} & 5.11 & \rho a_2 & 8.74 & K^*\text{K(1460)} & 7.63 \\
 \rho b_1\text{(2P)} & 4.15 & \rho a_1\text{(1640)} & 7.82 & KK^*\text{(3S)} & 7.39 \\
 \pi a_0\text{(5P)} & 3.9 & \rho a_1 & 7.63 & K_1K_2{}^* & 6.32 \\
 \text{$\rho $(1900)$\rho $} & 3.73 & \rho a_2\text{(1950)} & 6.76 & KK^*\text{(4S)} & 5.81 \\
 \text{$\pi \omega $(5S)} & 3.49 & \text{$\pi \rho $(5D)} & 4.25 & K^*K^* & 5.48 \\
 \rho b_1 & 3.29 & \rho \pi _2\text{(1880)} & 4.02 & K^*\text{K(1630)} & 4.44 \\
 \text{$\omega $(1960)$\pi $} & 3.27 & \text{$\pi \rho $(4D)} & 3.85 & K_4\text{(2045)K} & 4.38 \\
 \pi \omega _3\text{(4D)} & 3.15 & \rho \pi  & 3.38 & K^*K^*\text{(3S)} & 3.91 \\
 \omega a_2\text{(1700)} & 3.04 & \text{$\rho \pi $(1300)} & 3.25 & {K_1^\prime}K^* & 3.87 \\
 \rho f_2\text{(2P)} & 3.02 & \rho \pi _2 & 3.1 & K^*K_2\text{(1820)} & 3.65 \\
 \pi a_0\text{(4P)} & 3.01 & \text{$\pi \rho $(6S)} & 2.98 & K_1K^*\text{(1410)} & 3.63 \\
 \rho f_2 & 2.99 & \omega f_2 & 2.98 & K^*\text{(1410)}K_2{}^* & 3.42 \\
 \omega a_2 & 2.91 & \omega f_2\text{(2P)} & 2.91 & K^*K_2{}^*\text{(1980)} & 3.4 \\
 \rho b_1\text{(1960)} & 2.87 & \omega f_1\text{(2P)} & 2.64 & KK_3{}^*\text{(3D)} & 3.39 \\
 \pi a_4\text{(2255)} & 2.85 & \text{$\rho $(1700)$\pi $} & 2.4 & K_1K^* & 3.29 \\
 \rho f_1\text{(2P)} & 2.69 & \omega f_1 & 2.38 & K^*K_1'\text{(3P)} & 3.01 \\
 \omega a_1\text{(1640)} & 2.67 & \omega f_2\text{(1950)} & 2.2 & KK_2{}^*\text{(4P)} & 2.57 \\
 \rho \rho _2\text{(1D)} & 2.67 & h_1f_2 & 2.17 & K^*K_2{}^* & 2.55 \\
 \omega a_1 & 2.57 & \text{$\pi \rho $(3D)} & 2.09 & {K_1^\prime}K^*\text{(1410)} & 2.31 \\
 \rho f_2\text{(1950)} & 2.47 & \text{$\rho $(2000)$\pi $} & 1.9 & K_1K_0\text{(1430)} & 2.08 \\
 \rho f_1 & 2.4 & h_1f_0\text{(1500)} & 1.87 & K^*K_1'\text{(2P)} & 1.9 \\
 \pi a_0\text{(3P)} & 2.37 & b_1a_2 & 1.8 & KK_4{}^*\text{(2F)} & 1.84 \\
 \pi a_4\text{(3F)} & 2.02 & \pi \rho _3\text{(1690)} & 1.65 & K^*\text{(1410)K(1630)} & 1.81 \\
 \omega a_2\text{(1950)} & 1.99 & K^*\text{K(1830)} & 1.37 & KK_1{}^*\text{(3D)} & 1.7 \\
 a_1a_1 & 1.89 & \text{$\rho \pi $(1800)} & 1.32 & K_3{}^*\text{(1780)}K^* & 1.61 \\
 \rho \rho _3\text{(1690)} & 1.83 & - & - & KK^*\text{(5S)} & 1.58 \\
 \pi a_1\text{(1640)} & 1.51 & - & - & K^*\text{(1410)}K^*\text{(1410)} & 1.55 \\
 h_1a_2 & 1.47 & - & - & {K_1^\prime}K_1 & 1.38 \\
 \text{$\pi \omega $(5D)} & 1.42 & - & - & K^*K_2'\text{(2D)} & 1.36 \\
 \text{$\pi \omega $(4D)} & 1.29 & - & - &{K_1^\prime}K_2{}^* & 1.33 \\
 \rho \rho  & 1.28 & - & - & K^*K_1\text{(1650)} & 1.29 \\
 \omega \pi _2\text{(1880)} & 1.19 & - & - & KK_2{}^*\text{(2F)} & 1.28 \\
 \pi a_2 & 1.18 & - & - & KK_0{}^*\text{(3P)} & 1.24 \\
 f_2b_1 & 1.17 & - & - & f_1\text{(1420)$\phi $} & 1.14 \\
 K^*\text{K(1830)} & 1.12 & - & - & K^*K_2\text{(1770)} & 1.12 \\
 \pi a_1\text{(3P)} & 1.11 & - & - & \text{K(1460)}K_2{}^* & 1.12 \\
 \pi a_2\text{(3F)} & 1.11 & - & - & K^*\text{(1680)}K^* & 1.1 \\
 \omega \pi  & 1.09 & - & - & - & - \\
 \text{$\omega \pi $(1300)} & 1.01 & - & - & - & - \\
 \hline
  \hline
\end{array}\]
\end{table*}

\section{CONCLUSION}\label{sec4}
In this work, we have systematically studied the mass spectra and the OZI-allowed two-body strong decay behaviors of the newly observed $X(1910)$, $X(2300)$, and $b_1$, $h_1$ as well as $h_1^\prime$ states. 
We adopt new fitted parameters of the MGI model. Our findings can be summarized as follows:

 \begin{enumerate}

 \item{The newly observed state $X(1910)$ is a good candidate for the $s\bar{s}$ $h_1^\prime(2^1P_1)$ state. Based on the Regge trajectory and MGI model results, the $X(2300)$ can be identified as an $s\bar{s}$ $3^1P_1$ state within the mass spectrum. However, the decay width of the $X(2300)$ is too low compared to its experimental total width as the $h_1^\prime(3^1P_1)$ under the current theoretical framework.   }

 \item{The masses of $b_1(2P)$, $b_1(5P)$, $b_1(6P)$, and $b_1(7P)$, predicted by us, are about 1700 MeV, 2530 MeV, 2700 MeV, and 2825 MeV, respectively. The total widths are about 395 MeV, 390 MeV, 280 MeV, and 200 MeV for $b_1(2P)$, $b_1(5P)$, $b_1(6P)$, and $b_1(7P)$, respectively. The total and partial widths can be seen in Tables \ref{1pdecay}, \ref{2pdecay}, \ref{3pdecay}, \ref{4pdecay}, \ref{5pdecay}, \ref{6pdecay} and \ref{7pdecay}.}

 \item{The $h_1(1170)$, $h_1(1595)$, $h_1(1965)$ and $h_1(2215)$ can be identified as the $1^1P_1$, $2^1P_1$, $3^1P_1$  and $4^1P_1$, respectively. The masses and decay modes of $h_1(5P)$, $h_1(6P)$, and $h_1(7P)$ are presented in Tables \ref{mass}, \ref{5pdecay}, \ref{6pdecay} and \ref{7pdecay}.}
 
 \item{As for the excited states of $h_1(1415)$, the position of $h_1^\prime(3P)$ is currently open and awaiting for the best candidate. The masses and decay information of $h_1^\prime(4P)$, $h_1^\prime(5P)$, $h_1^\prime(6P)$, and $h_1^\prime(7P)$ are shown in Tables \ref{mass}, \ref{4pdecay}, \ref{5pdecay}, \ref{6pdecay} and \ref{7pdecay}. }

\end{enumerate}

We look forward to upcoming experimental studies that will play a crucial role in clarifying the nature of the newly observed light hadron family members with $J^{PC}=1^{+-}$, as well as in validating or exploring these theoretical predictions.

\begin{acknowledgments}
This work is supported by the National Natural Science Foundation of China under Grants  No.~12235018, No.~11975165, No.~11965016, and No.~12247101, and by the Natural Science Foundation of Qinghai Province under Grant No. 2022-ZJ-939Q, the Fundamental Research Funds for the Central Universities (Grant No. lzujbky-2024-jdzx06).
\end{acknowledgments}

\newpage
\bibliographystyle{apsrev4-1}
\bibliography{hepref}
\end{document}